\documentclass[pre,aps,amsmath,amssymb,eqsecnum,floatfix,twocolumn]{revtex4-1}

\usepackage{ctex}
\usepackage{graphicx}
\usepackage{euscript}
\usepackage{oldgerm}
\usepackage{subfigure}
\usepackage{color}
\usepackage{txfonts}
\usepackage{amssymb}
\usepackage{amsmath}

\graphicspath{{./}{../circle_swimmers_2D_v5/}}

\begin{document}

\title{Self-Organized Vortices of Circling Self-Propelled Particles and Curved Active Flagella}

\author{Yingzi Yang}
\author{Feng Qiu}
\affiliation{The State Key Laboratory of Molecular Engineering of Polymers,
Department of Macromolecular Science, Fudan University, Shanghai 200433, China}

\author{Gerhard Gompper}
\affiliation{Theoretical Soft Matter and Biophysics, Institute of
Complex Systems and Institute for Advanced Simulation,
Forschungszentrum J{\"u}lich, D-52425 J{\"u}lich, Germany}

\begin{abstract}
Self-propelled point-like particles move along circular trajectories when their
translocation velocity is constant and the angular velocity related to their
orientation vector is also constant. We investigate the collective behavior of
ensembles of such
circle swimmers by Brownian dynamics simulations. If the particles interact
via a ``velocity-trajectory coordination" rule within neighboring particles,
a self-organized vortex pattern emerges. This vortex pattern is characterized
by its particle-density correlation function $G_\rho$, the density correlation
function $G_c$ of trajectory centers, and an order parameter $S$ representing
the degree of the aggregation of the particles. Here, we systematically
vary the system parameters, such as the particle density and the interaction
range, in order to reveal the transition of the system from
a light-vortex-dominated to heavy-vortex-dominated state, where vortices contain
mainly a single and many self-propelled particles, respectively.
We also study a semi-dilute solution of curved, sinusoidal-beating flagella,
as an example of circling self-propelled particles with explicit propulsion
mechanism and excluded-volume interactions. Our simulation results are compared
with previous experimental results for the vortices in sea-urchin sperm solutions
near a wall. The properties of the vortices in simulations and experiments are
found to agree quantitatively.
\end{abstract}

\maketitle

\section{Introduction}
\label{sec:intro}

Systems of self-propelled particles (SPPs), which consume energy to maintain a
persistent non-Brownian motion, exhibit an abundance of fascinating non-equilibrium
collective behaviors --- such as swarming, swirling, and clustering \cite{Vicsek2012}.
Examples are found in very different areas of physics and biology, and range from
actin filaments and microtubules in motility assays
\cite{Schaller2010,Butt2010,Sumino2012} through metallic
nano-rods \cite{Paxton2004,Dhar2006} to flocking of birds \cite{Bajec2009,Hayakawa2010}
and groups of people \cite{Faria2010,Moussaid2011}. In all of these examples, SPPs
employ different propulsion mechanisms, and also interact with each other
in different ways, including direct physical contact \cite{Peruani2006,Yang2010},
chemotaxis \cite{Czirok1996}, hydrodynamic interactions
\cite{Hernandez-Ortiz2005,Pooley2007,Yang2008}, and restricted visual contact
\cite{Strombom2011}. However, the collective behavior of SPPs
systems is surprisingly similar, displaying phenomena such as giant density
fluctuations with anomalously slow relaxation \cite{Simha2002,Narayan2007,Aranson2008},
order-disorder phase transition with increasing noise and decreasing density
\cite{Vicsek1995,Baglietto2009,Kulinskii2009,Chate2008,ZhiXin2008,Ginelli2010}, etc.

Since the simplest model of interacting SPPs was introduced by Vicsek et al.~\cite{Vicsek1995},
nowadays widely referred as the ``Vicsek model"
\cite{Baglietto2009,Baglietto2009b,Kulinskii2009,Chate2008,ZhiXin2008,Ginelli2010},
collective behavior of SPP systems has attracted much interest \cite{Vicsek2012}.
In the Vicsek model, point particles moving with constant velocity align their
directions of motion with the average direction of other particles in a prescribed
interaction range, while internal or external noise is taken into account by adding a
random increment to their orientation vectors. By a variation of
the system parameters, e.g. the particle density and the strength of the perturbation,
such a simple system has been found to undergo an order-disorder phase transition,
whose nature (first or second order) is related to the particle velocity and the way
the perturbation is introduced \cite{Baglietto2009,Mishra2010,Ihle2011}. According to
these studies, the initial conditions and settings of the simulation, the type of
interaction and the boundary conditions play an important role in the formation of
certain collective patterns of motion \cite{Vicsek2012}.

A special, but not rare class of patterns of collective motion in SPPs systems are
swirls or vortices, in which a group of particles circle around a common center. In
experiments, swirls and vortices emerge both in non-living particle systems (vertically
vibrated granular rods \cite{Blair2003}, motility assays of actin filaments
\cite{Schaller2010} or microtubules \cite{Sumino2012}) and in systems of
living micro-organisms (bacteria colonies
\cite{Czirok1996}, zoo-plankton under optical stimulus \cite{Ordermann2003}, sea-urchin
sperm trapped near a substrate \cite{Riedel2005}). In simulations,
the emergence of swirls and vortices was found to depend on the initial conditions and
the model settings, such as a circular boundary \cite{Grossman2008, Kudrolli2008},
alignment with a ``blind angle" of interaction behind each agent \cite{Strombom2011},
a harmonic attractive pair potential with a noise above a critical value \cite{Erdmann2005},
and hydrodynamic interactions \cite{Hernandez-Ortiz2005}. In most of the computational
models exhibiting vortices, a single SPP is either assumed to perform a random
walk \cite{Ordermann2003}, or to move with constant magnitude of velocity
\cite{Czirok1996,Grossman2008,Strombom2011,Erdmann2005}.

In this paper, we consider a class of SPPs which move --- in the absence of noise ---
along {\em curved trajectories} rather than straight lines, which we call circle SPPs.
The driving force of a circle SPP does not coincides with its propagation direction, so
that its trajectory is a cycloid in three spatial dimensions or a circle in two dimensions
\cite{Teeffelen2008,Teeffelen2009,Ledesma-Aguilar2012}. Artificial circle swimmers can be
constructed by introducing a tilted or bend structure to catalytically driven colloidal rods
\cite{Dhar2006}, giving thermophoretic colloidal swimmers a L-shape \cite{Kuemmel2013},
or by designing micro-machines of connected beads which are moved
relative to each other in a time-irreversible multi-step cycle \cite{Dreyfus2005}.
Examples of circle swimmers in living systems include certain bacteria \cite{Lauga2006}
and spermatozoa \cite{Moore1995,Riedel2005,Elgeti2010} --- when the micro-organisms are
attracted or confined in their motion
to a surface or wall. The most carefully analyzed experiment on the collective motion of
circle SPPs might be the self-organized vortices of sea-urchin sperm \cite{Riedel2005}.
The 50-$\mu$m-long sperm circle clockwise when they gather at a substrate.
With increasing surface density of sperm, the system exhibits a transition from a
disordered state of randomly distributed sperm to a self-organized vortex-array state
with local hexagonal order, in which each vortex consists of several sperm.
In Ref.~\onlinecite{Riedel2005}, the emergence of this structure is attributed to the
hydrodynamic interactions between individual sperm and between sperm vortices. However,
simulations suggest that the steric interactions between rod-like SPPs
\cite{Peruani2006,Yang2010} and sinusoidal beating flagella \cite{Yang2010} strongly
contribute to the collective motion and aggregation.

While the morphology and properties of swirls and vortices in straight-trajectory SPPs
systems have been studied intensively in recent years
\cite{Czirok1996,Grossman2008, Kudrolli2008,Strombom2011,Erdmann2005,Hernandez-Ortiz2005,Voituriez2006},
much less is known theoretically about the collective motion of circle SPPs
\cite{Sumino2012,Kaiser2013}. The
understanding of the effect of the spontaneous circular trajectory on the collective
behavior of SPPs is essential for the understanding of many biological phenomena and
for the design of microscopic machines. In addition to the parameters of SPPs (like
velocity, particle density, interaction range, and noise amplitude), the circle SPP
systems have at least the spontaneous curvature of the particle trajectory as an extra
parameter. The larger parameter space suggests a more complex behavior of these systems.
In order to study such systems, highly simplified models are often very useful, as has been
demonstrated by the success of the Vicsek model.
An example for circle swimmers is the simplified model of Ref.~\onlinecite{Riedel2005},
designed to interpret the formation of vortex arrays of sea urchin sperm. In this model,
each sperm is described by a point particle at its trajectory center, with a pairwise
short-range attraction arising from hydrodynamic forces, and a longer range repulsion due
to steric or hydrodynamic origin. Inspired by this model, we construct a model of circle
SPPs by point-like particles with a constant propagation velocity, where particles interact
through a ``velocity-trajectory coordination'', which takes the trajectory centers of
neighbor particles into account. This algorithm of interaction is different from both the
``trajectory-center coordination" of Ref.~\onlinecite{Riedel2005} and the
``velocity coordination" of the Vicsek model \cite{Vicsek1995} (which averages the
velocity of neighboring particles). We will analyze this model in detail, and show that
it describes vortex formation and the evolution of a stationary vortex pattern.

In addition to the point-like circle SPPs model, we also study curved, sinusoidal-beating
flagella.  Sinusoidal beating of a filament or elongated body is a common self-propulsion
mechanism in biological systems with low-Reynolds-number hydrodynamics, e.g. nematodes
\cite{Gray1964} and sperm of higher animals \cite{Gray1955,Riedel-Kruse2007}. The sinusoidal wave
propagates from one end to the other on the filament-like body and pushes the surrounding
fluid backwards to generate a forward force. Thus, the cell or organism gains velocity
in opposite direction of wave propagation.  As indicated by experiments of biological
systems \cite{Gray1964,Moore2002,Cisneros2011,Peruani2012} and simulations of model
systems \cite{Yang2010}, the sinusoidally undulating motion of the  body does not destroy
the general collective behaviors of rod-like SPPs. Our flagellum model is
coarse-grained as particles connected by harmonic springs, and the hydrodynamics is either
approximated by anisotropic friction, or calculated by using Multi-Particle Collision
dynamics (MPC), a mesoscopic particle-based simulation approach \cite{Kapral2008,Gompper2009}.
By introducing a non-zero average curvature in the beating plane, the undulating flagellum
traces out a circular trajectory in two dimensions, reminiscent the trajectory of sea urchin
sperm at a substrate \cite{Kaupp2003,Riedel2005}. The simplicity of the model allows us
to analyze the contribution of volume exclusion and hydrodynamic interaction separately.
The study provides insight into the effect of flagellar properties, such as
frequency distribution and spontaneous curvature, on the collective motion and the
emerging vortex patterns.

The paper is organized as follows. Section \ref{sec:models} gives a brief description of our models
and simulation methods. In Sec.~\ref{sec:SPP}, we analyze the collective motion of
point-like circle SPPs systems. Then, we study the collective motion of curved, sinusoidally beating
flagella, and compare the results with circle SPPs models and sea-urchin sperm experiments in
Sec.~\ref{sec:flagella}. The results are summarized in Sec.~\ref{sec:summary}.

\section{Models}
\label{sec:models}

\subsection{Circle self-propelled particles}

\begin{figure}
\includegraphics[width=7cm]{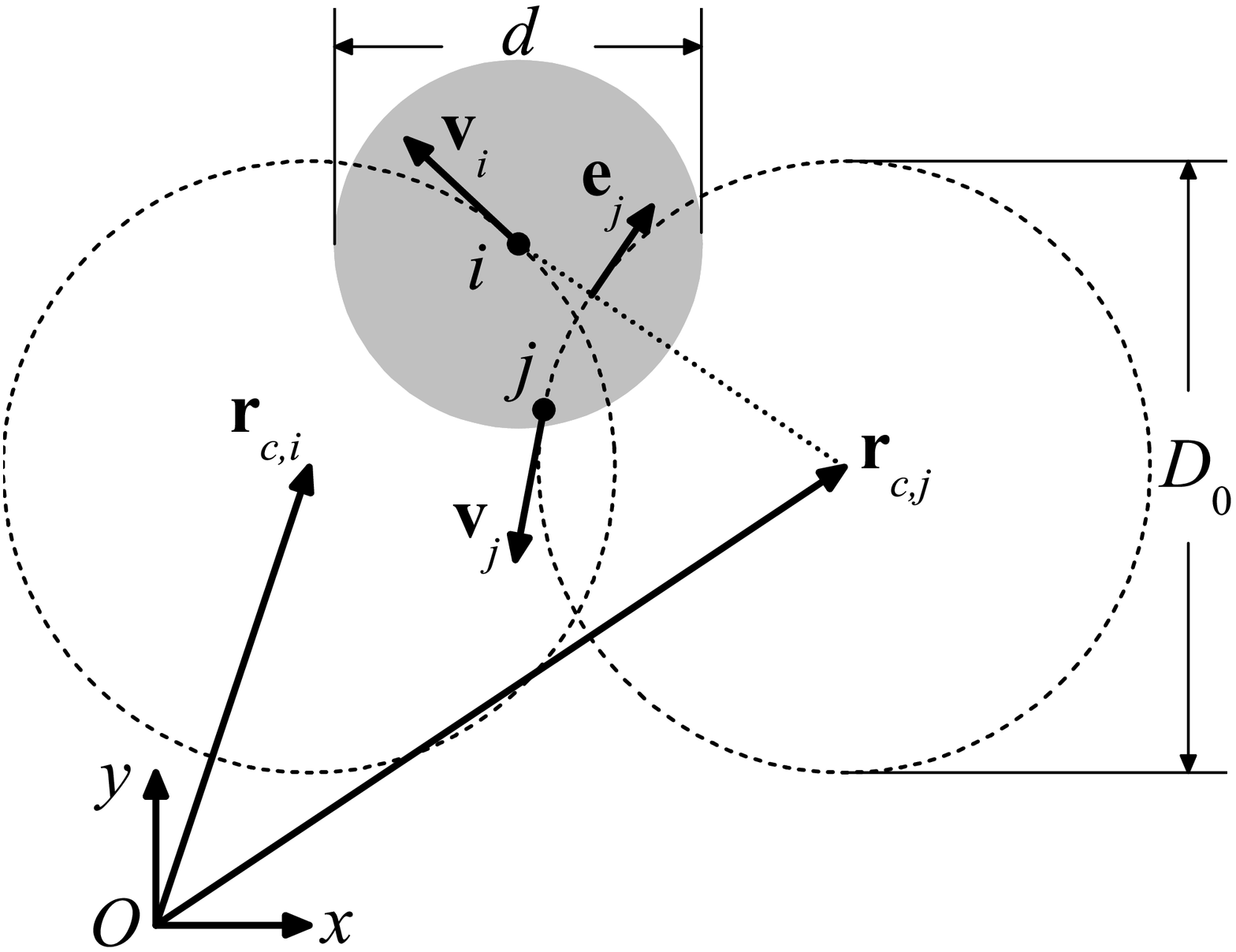}
\caption{Schematic of the interaction of particles $i$ and $j$. The dashed lines denote the
spontaneous circular trajectories of the $i$-th and $j$-th particles. The gray area displays
the neighbor region of particle $i$. ${\bf e}_j$ is the tangent unit vector to the trajectory
of particle $j$ with ${\bf v}_i\cdot {\bf e}_j \ge 0$.}
\label{fig:circularswimmermodel}
\end{figure}

We consider $N$ point-like particles moving in a two-dimensional system of size $L_x \times L_y$.
The number density of the particles is $\rho_0 = N/L_x L_y$. Each particle has a spontaneous
circular trajectory of diameter $D_0$, which is traversed in counter-clockwise direction,
and a circular interaction region of diameter $d$, as illustrated in
Fig.~\ref{fig:circularswimmermodel}. At time $t$, the $i$-th particle has position ${\bf r}_i$,
velocity ${\bf v}_i$, and trajectory center position ${\bf r}_{c,i}$ of its spontaneous circle
trajectory determined by
\begin{equation}
 {\bf r}_{c,i}(t)={\bf r}_i(t) + \frac{D_0}{2}\mathfrak R(\frac{\pi}{2})\frac{{\bf v}_i(t)}{|{\bf v}_i(t)|},
\end{equation}
where $\mathfrak R(\theta)$ is a rotation matrix which rotates a vector in two dimensions
counter-clockwise through an angle $\theta$. Note that ${\bf r}_{c,i}(t)$ is obtained from the
instantaneous velocity and position, rather than from the full trajectory.

The equation of motion for the $i$-th particles is then given by
\begin{equation}
{\bf r}_i(t+\Delta t)={\bf r}_i(t)+{\bf v}_i(t+\Delta t)\Delta t.
\label{eq:CSeq1}
\end{equation}
Motivated by the nematic alignment
of colliding rod-like particles \cite{Ginelli2010}, we define a majority rule for the
the re-direction of the velocity as
\begin{equation}
{\bf v}_i(t+\Delta t)=v_0\mathfrak R(\omega_0\Delta t)
\frac{\sum_{j(i)}{\bf e}_j}{|\sum_{j(i)}{\bf e}_j|},
\label{eq:CSeq2}
\end{equation}
where $\Delta t$ is the simulation time step, $v_0$ is constant magnitude of the velocity,
$\omega_0=\frac{2v_0}{D_0}$ is the angular velocity of the orientation of the
particle velocity which enforces a circular trajectory of diameter $D_0$ in dilute
suspension, and $\sum_{j(i)}$ is the sum of all particles within the interaction
region of $i$-th particle -- including $i$-th particle itself.
The tangential unit vector ${\bf e}_j$ is obtained at the intersection of the spontaneous
trajectory of $j$-th particle and the connect line between ${\bf r}_i$ and ${\bf r}_{c,j}$,
as illustrated in Fig.~\ref{fig:circularswimmermodel}. The direction of ${\bf e}_j$ is
chosen to make ${\bf v}_i\cdot {\bf e}_j \geq 0$.
With such an interaction rule, a particle tends to move parallel or
anti-parallel to the trajectories of its neighbors. Therefore, the result of the interaction
is a local nematic alignment of trajectories.

In contrast to the Vicsek model \cite{Vicsek1995, Vicsek2012}, the change of the
velocity direction in our model is determined not by a polar alignment of velocities
(velocity coordination), but rather by a nematic alignment of local trajectories, which
are determined by the instantaneous trajectory centers ${\bf r}_{c,j}$ of the neighbor
particles (velocity-trajectory coordination).
Here, the choice of the alignment rule strongly depends on the interaction range relative
to the preferred swimming radius. Clearly, a majority rule can only generate alignment if
the number of particles in the interaction range is sufficiently large. This is
very difficult to achieve for $d \ll D_0$ when particles move collective in circular
vortices. Therefore, we choose $d \simeq D_0$.
Our simulations show that in this case velocity coordination as in the
Vicsek model does not give rise to vortex patterns in a circle SPP system,
because neighboring particles align their directions of motion but do not
tend to generate vortices with several particles circling around a common center.

Note that we omit noise terms in Eqs.~(\ref{eq:CSeq1}) and (\ref{eq:CSeq2}), so
that our model is deterministic -- once the initial state is given. This is a reasonable
description of microswimmers which are not too small, such as sperm.

In the simulation, we use the diameter of spontaneous trajectory
$D_0$ as length unit, $\Delta t$ as time unit, and $v_0=0.025 D_0/\Delta t$.
This implies a period of rotation of an interaction-free particle of $T_0=\pi D_0/v_0$.
The interaction range $d$ is varied below, but is typically
on the order of $D_0$ (as explained above).
The simulation box size is $L_x=L_y=20D_0$ or $40D_0$.
The simulation time for each system is $T=5\times 10^5\Delta t \approx 4.0\times 10^3T_0$.

\subsection{Circle-swimming flagella}

We construct a two-dimensional flagellum model \cite{Yang2008} from a linear sequence of $N_f$ beads, which are
connected by harmonic springs with rest length $l_0$. The local spontaneous curvature of such
a filament is a function of time $t$ and position $x$ along the flagellar contour, measured from the first bead,
\begin{equation}
c(x,t)=c_0+A \sin(-2\pi ft+qx+\varphi_0),
\label{eq:beat}
\end{equation}
where $f$ is the beating frequency, $q$ is the wave number, $\varphi_0$ is the initial phase
chosen independently from a uniform distribution on the interval $[0,2\pi)$ for each flagellum,
and $c_0$ is the average curvature. We denote a flagellum with non-zero $c_0$ a curved flagellum.
The constant $A$ controls the strength of beating, which is related to
the amplitude of the shape undulations. Equation~(\ref{eq:beat}) generates a traveling wave, which
propagates from the front to the end of the flagellum. The curvature elasticity of the
flagellum is determined by a bending potential, which depends on the deviations of the
angles between neighbor bonds from their preferred value $c(x,t)l_0$. Therefore, the
elastic energy of flagellum $j$ is given by
\begin{equation}
  V_j = \sum_{i=1}^{N_f-1}\frac{1}{2}\frac{k}{l_0^2}\biggl[|{\bf R}_{i,j}|-l_0\biggr]^2
   + \sum_{i=1}^{N_f-2}\frac{1}{2}\frac{\kappa}{l_0^3}\biggl[{\bf R}_{i+1,j} -
                                   \mathfrak R(c_j(x_i,t)l_0){\bf R}_{i,j}\biggr]^2.
\label{eq:potential}
\end{equation}
Here, ${\bf R}_{i,j}={\bf r}_{i+1,j}-{\bf r}_{i,j}$ are bond vectors, and
${\bf r}_{i,j}$ denotes the position of the $i$-th bead of the $j$th flagellum.
$\mathfrak R(c_jl_0)$ is an operator rotating a two-dimensional vector counter-clockwise
by an angle $c_j(x_i,t)l_0$. $k$ is the spring constant and
$\kappa$ the bending rigidity \cite{parameters_flagellum}. The volume exclusion of beads
on different flagella is taken into account by a purely repulsive, truncated and
shifted Lennard-Jones potential
\begin{equation}
V_{LJ}(r)=\left\{\begin{array}{l l}
  4\epsilon\biggl[ \biggl(\frac{l_0}{r}\biggr)^{12}-\biggl(\frac{l_0}{r}\biggr)^6+\frac{1}{4}\biggr],
                                                                   & r\leq2^{1/6}l_0\\
  0, & \text{otherwise }   \\
\end{array}\right. \ ,
\label{eq:LJpotenttial}
\end{equation}
where $r$ is the distance between two beads belonging to different flagella.

The swimming of both nematodes and sperm is determined by low-Reynolds-number hydrodynamics,
where viscous forces dominate over inertial forces. In this regime, the dynamics of a
rod can often be well described by resistive-force theory \cite{Gray1955},
in which hydrodynamic interactions between different segments of a rod are
approximated by an anisotropic friction (AF), so that
${\bf f}_{\parallel}=-\gamma_{\parallel}{\bf v}_{\parallel}l_0$, ${\bf f}_{\perp}
=-\gamma_{\perp}{\bf v}_{\perp}l_0$,
where ${\bf v}_{\parallel}$ and ${\bf v}_{\perp}$ are the velocity components of a flagellum
segment on the directions parallel and perpendicular to the local tangent vector, respectively.
$\gamma_B$ is the friction coefficient for a segment of unit length, with $B=\parallel$ or
$\perp$ \cite{parameters_AF}.
Hydrodynamic interactions are not included in simulation employing AF, so that
volume exclusion is the only interaction between the flagella in this case. Considering the
relatively large sizes of flagella and nematodes of several 10 $\mu$m and a few millimeters,
respectively, we neglect thermal fluctuations in our AF simulations.

In simulations with full hydrodynamics, the flagellum model is embedded in a
two-dimensional fluid, and the time evolution of the system is carried out by employing
a hybrid molecular dynamics approach --- multi-particle collision dynamics (MPC) for
the fluid \cite{Kapral2008,Gompper2009}
in combination with molecular dynamics (MD) for the flagellum --- as follows.
During the streaming step, the point-like particles of the MPC fluid move ballistically during
a time interval $\Delta\tau$, while Newton's equations of motion for the flagellum particles
are integrated by the Verlet algorithm with a time step of $\Delta \tau'=0.02 \Delta \tau$.
In the subsequent collision step, the flagella exchange momentum with the neighboring fluid
particles.
The collisions are performed by sorting all fluid and flagellum particles into the cells
of a cubic lattice (with lattice constant $a$), which are
labeled by an index $\xi$. A rotation operator $\mathfrak R_\xi(\alpha)$ is assigned to
each box. If ${\bf v}_{c,\xi}$ is the pre-collisional center-of-mass velocity of all particles in
box $\xi$, the post-collisional velocity ${\bf v}_i'$ of a particle $i$ in the box is given by
${\bf v}_i'={\bf v}_{c,\xi}+\mathfrak R_{\xi}(\alpha)\biggl({\bf v}_i-{\bf v}_{c,\xi}\biggr)$,
where ${\bf v}_i$ is its pre-collision velocity. This simple collision rule
conserves mass, momentum and energy, which guarantees the emergence of Navier-Stokes
hydrodynamics on length scales larger than $a$. Since the beating
flagella perpetually inject energy into the fluid, we apply a local thermostat
after each collision step.

\begin{figure}
\includegraphics[width=7cm]{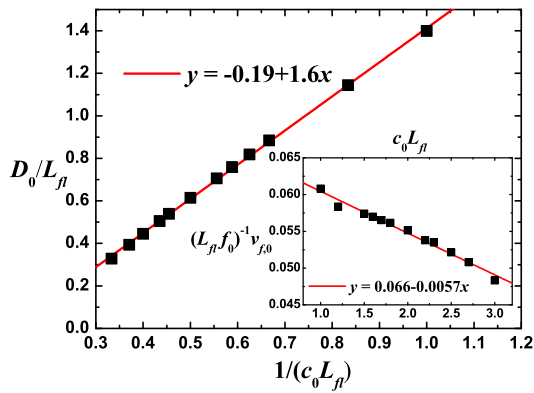}
\caption{(Color online) The diameter $D_0$ (black solid squares) of the circular trajectory of
a single flagellum as a function of $(c_0L_{fl})^{-1}$ in AF simulations. The red line is a linear fit
to the data with slope of 80. The inset is the dimensionless velocity $v_0/(f_0L_{fl})$ of a single flagellum
(black solid squares) as a function of $c_0L_{fl}$. The red line is the linear fit to the data.
In both (a) and (b), the beating frequency of the flagellum is $f_0=\tau_0^{-1}/120 $.
}
\label{fig:Graph_bend_basic}
\end{figure}

The simulation system contains $N$ flagella of length $L_{fl}=N_f l_0=50a$ in a simulation box of size
$L_x\times L_y$, where $a$ is size of a MPC collision box.
The wave number $q=4\pi/(N_f l_0)$ implies the presence of two complete sinusoidal
waves on the flagellum.
The number density of flagellum is
$\rho_0=N/L_x L_y$. Periodic boundary conditions are employed.
During the simulation, each flagellum has a constant frequency $f$ chosen
from a Gaussian distribution with the average $f_0=\tau_0^{-1}/120$ and the variance
$\langle(f-f_0)^2\rangle=\sigma^2f_0^2$, where $\sigma$ is a dimensionless number characterizing
the width of the frequency distribution. When the average spontaneous curvature $c_0$ is positive,
the flagellum is curved and prefers a clockwise circular trajectory.
The trajectory diameter $D_0$ depends approximately linearly on $(c_0L_{fl})^{-1}$ in AF simulations,
and the center-of-mass velocity $v_{f,0}$ of the flagellum decreases approximately linearly
with $c_0L_{fl}$ from its value for $c_0=0$, as shown in
Fig.~\ref{fig:Graph_bend_basic}.
Therefore, for a flagellum beating with the average frequency $f_0$, it takes
$L_{fl}f_0/v_{f,0}\approx(0.066-0.0057c_0 L_{fl})^{-1}$ beats to move a one body length.
Equivalently, the swimming velocity is $v_{f,0} \approx (0.066-0.0057c_0 L_{fl}) L_{fl} f_0$.
The beats number for a flagellum to complete a full circle trajectory is
$T_0f_0=\pi D_0f_0/v_{f,0}\approx\pi[-0.19+1.6/(c_0L_{fl})]/[0.066- 0.0057 c_0L_{fl}]$. For example,
in our AF simulations, the most strongly curved flagellum, with $c_0 L_{fl}=3$,
takes $T_0f_0\approx 20$ beats to complete a circle, while the least curved flagellum,
with $c_0 L_{fl}=1$, takes $T_0f_0\approx 75$ beats.
Each simulation runs for at least $2500/f_0$.
Therefore, the flagella beat at least 2500 times on average in a simulation run,
and complete about 35 to 110 full circles.
The other parameters of our flagellum simulations are listed in Ref.~\onlinecite{parameters_MPC}.

The collision of two particles with elongated structure in a viscous fluid
environment results in a cooperated motion that the particles move together with close
packing. If the elongated particles are straight, the nematic interaction results in velocity
alignment of neighbor particles and cluster formation \cite{Yang2010,Peruani2006,Ginelli2010}.
When two elongated and curved flagella encounter each other, they tend to get close and synchronize
their configuration via hydrodynamic interactions as well as volume exclusion \cite{Yang2008,Yang2010}
and form an effective ``extended flagellum" with the same average curvature. The circular
motion of such an extended flagellum around a center is equivalent to the motion of two flagella
around the same center, the first step towards the formation of a vortex.

\section{Collective Motion of Circle SPPs}
\label{sec:SPP}

\begin{figure}
\includegraphics[width=7cm]{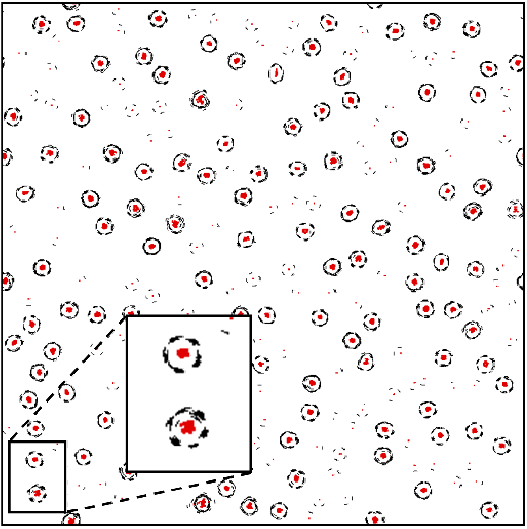}
\caption{(Color online) A snapshot of vortices in a circle SPPs system at the end of the
simulation time $t=4\times10^3T_0$. Each black dash is the trajectory of a particle in a
time interval $\Delta T = 10\Delta t \approx 0.08 T_0$, and the red dots are the corresponding
trajectory centers. The inset shows a magnification of the area at the bottom left corner where
two vortices consists of groups of circle SPPs with equal distance between neighboring groups.
The number of particle groups of them are five and six, respectively. The parameters are
$L_x=L_y=40D_0$, $\rho_0D_0^2 = 4$, and $d=1.3D_0$. }
\label{fig:particle_snapshot}
\end{figure}

$N$ circle SPPs are distributed in the simulation box with random positions and random
velocity directions at the initial time $t=0$. Shortly after the start, vortices emerge,
which are formed by one or several particles circling around a common center,
as illustrated in Fig.~\ref{fig:particle_snapshot} and movies in the supplemental
material \cite{movie}. The
vortices distribute in space with no obvious long-range order, and displace vigorously if
the trajectory centers of the particles belonging to the same vortex do not coincide precisely
at the same point. When two vortices overlap or collide, particles are exchanged between
the vortices until they either fuse or separate from each other far enough. Therefore, the
vortex mass, defined as the number of circle SPPs forming it, changes during collisions.
Note that the system is deterministic due to the absence of noise, and a noise-induced
fission of vortices does not occur in the simulation.
We characterize the emergent vortex pattern of circle SPPs by the particle density-correlation
functions, the trajectory-center density-correlation functions, and the order parameter
reflecting the degree of the particle aggregation.
By systematically changing the particle density $\rho_0$ and the size of the interaction
region $d$, we then determine the dynamics state diagram of this system.

\subsection{Particle density-correlation functions and trajectory-center density-correlation
functions}
\label{sec:SPP_corr}

The correlation function of particle densities at time $t$ is given by
\begin{equation}
G_\rho(|{\bf r}-{\bf r}'|)=\rho_0^{-2}\langle\rho({\bf r},t)\cdot\rho({\bf r}',t)\rangle_t,
\end{equation}
and the correlation function of the trajectory-center density at time $t$ is
\begin{equation}
G_c(|{\bf r}-{\bf r}'|)=\rho_0^{-2}\langle\rho_c({\bf r},t)\cdot\rho_c({\bf r}',t)\rangle_t,
\end{equation}
where $\rho({\bf r},t)$ and $\rho_c({\bf r},t)$ are the number densities of the particles
and the trajectory centers at position ${\bf r}$ and time $t$, respectively. We consider times $t$
in the interval $[T_1, T_2]$, where $T_1 = 8\times 10^2T_0$ and $T_2=4\times 10^3T_0$.
Most of the systems have not yet reached a stationary state at $t=T_1$, so that the correlation
functions $G_\rho(x,t)$ and $G_c(x,t)$ still contain some averaging over different
structures and spatial arrangements of the vortices.

\begin{figure}
\includegraphics[width=7cm]{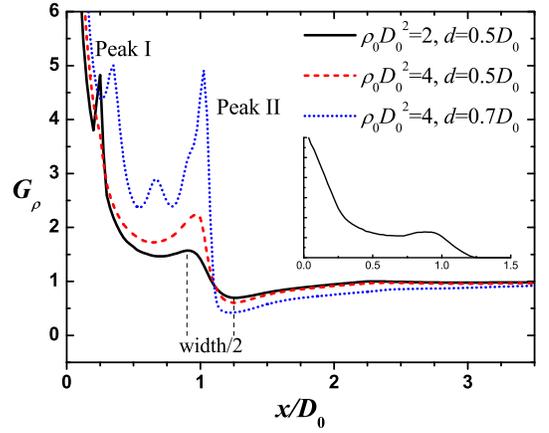}
\caption{(Color online) The examples of the particle-density correlation function $G_\rho$.
The inset is the density correlation function of an annular region of evenly distributed
particle density. The diameter of the inner circle of the annular region is $0.75 D_0$, and the
diameter of the outer circle is $1.25 D_0$. }
\label{fig:Graph_particle_Grho}
\end{figure}

\begin{figure}
\includegraphics[width=7cm]{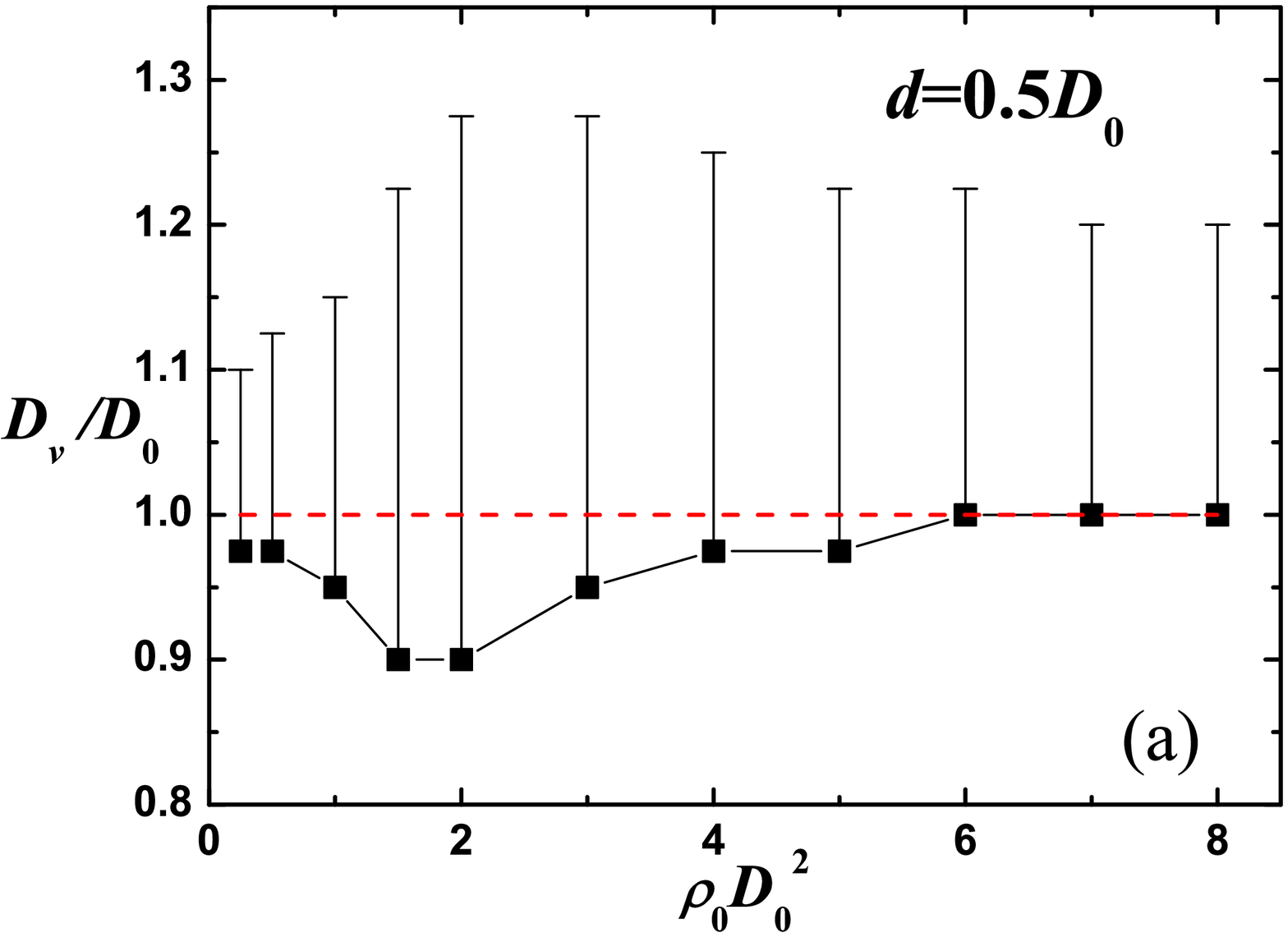}\\
\includegraphics[width=7.3cm]{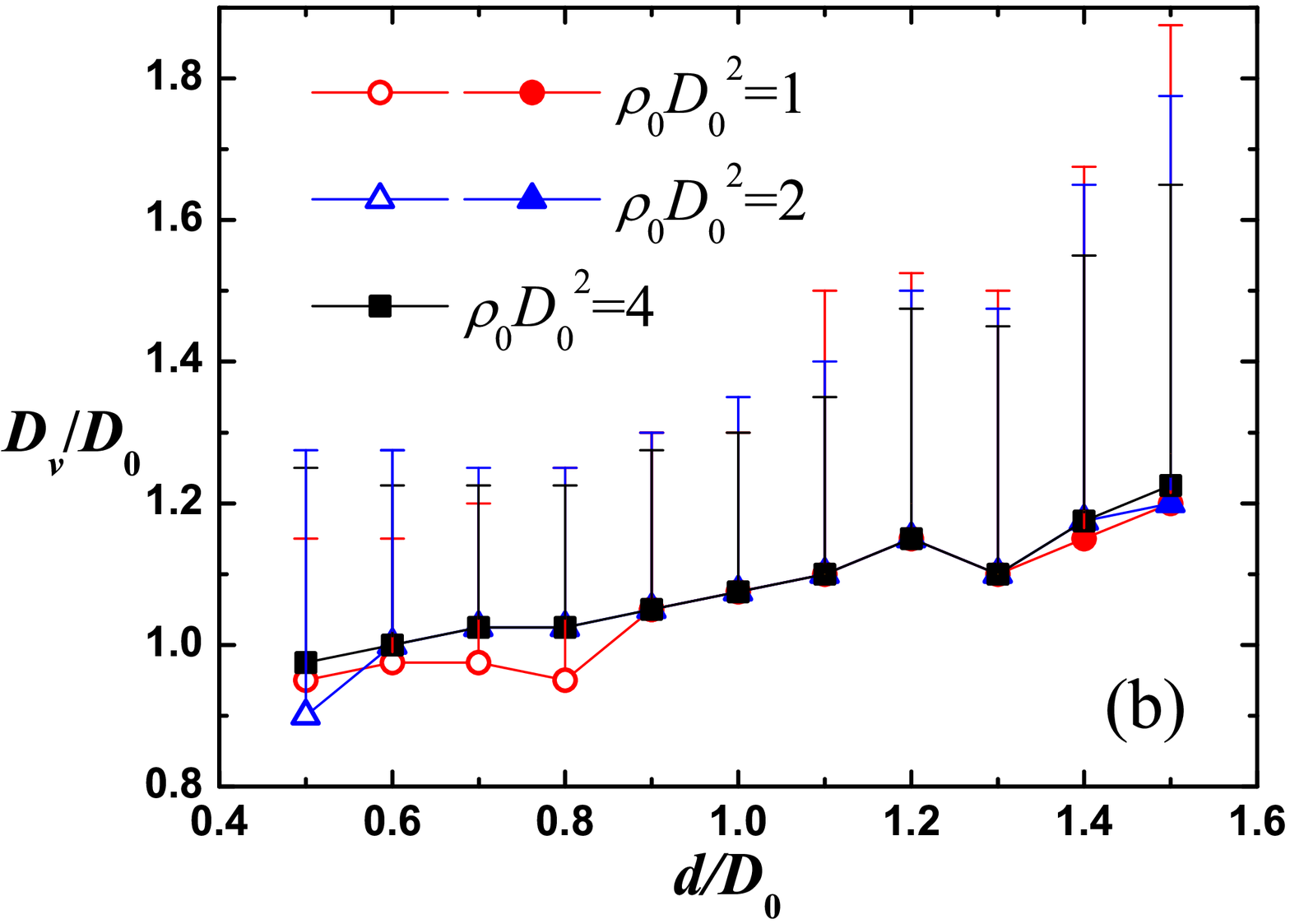}
\caption{(Color online) (a) The vortex diameter $D_v$ as a function of the particle density
$\rho_0D_0^2$. The red dashed line is the guide to the eye at value $D_v=D_0$.
(b) $D_v$ as a function of
the diameter of the interaction region $d$. The open symbols and the solid symbols represent
the light-vortex-dominated systems and the heavy-vortex-dominated systems, respectively.
The ``error bars" in (a) and (b) indicate the half width of peak II as defined in
Fig.~\ref{fig:Graph_particle_Grho}, and therefore characterize the width of the band of
particle trajectories within a vortex.
}
\label{fig:particle_Dv}
\end{figure}

The particle density-correlation function $G_\rho(x)$ mainly characterizes particle arrangement in a
vortex, in particular for $x<D_0+d/2$, as shown in Fig.~\ref{fig:Graph_particle_Grho}.
In the single-vortex region, $G_\rho(x)$ displays a pronounced spatial dependence.
There is a pronounced peak at $x\simeq d/2$, denoted as peak I, which arises from the
interaction range of our model. The displacement of a particle in a simulation
time step, $v_0\Delta t$, is much smaller than $d/2$; therefore, during the very short time
interval for two particles to adjust their motion and form a vortex, their relative distance
remains essentially constant. This distance only changes when a third particle comes into play.
If the mass of a vortex is high, the multi-particle interaction affects and disturbs the
distances of other particles in a vortex. Thus, a higher density $\rho_0$ depresses peak I
in Fig.~\ref{fig:Graph_particle_Grho}.
For example, when $d=0.5D_0$, doubling $\rho_0$ strongly depresses the height of peak I.
Similarly, increasing $d$ enhances peak I when $\rho_0$ is kept
constant. Therefore, a vortex is composed of several groups of particles with the constant
distance $d/2$ between
neighbor groups, as illustrated in Fig.~\ref{fig:particle_snapshot}. If the tendency to keep
the distance $d/2$ is strong for the particle groups, a second and third peak become evident,
see Fig.~\ref{fig:Graph_particle_Grho}. The position of the $n$-th peak
is approximately $x_n=D_v\sin (n\theta/2)$ where $\theta = 2\arcsin(d/2D_v)$ and $D_v$ is
the diameter of the vortex. Note that such periodic particle density modulation is a dynamic
temporary structure which breaks and reconstructs with time.

The peak of the vortex structure at $x\simeq D_0$, denoted as peak II, indicates the
diameter $D_v$ of a vortex. According to the definition of the density correlation function
$G_\rho$, this diameter is weighted by the mass of the vortices, so that the vortices consisting
of more particles contribute more to $D_v$. Thus, $D_v$ mainly reflects the diameter of the
``heavy" vortices, containing a large number of particles in the vortex. We define the
width of peak II as the distance between the peak position and the position of the subsequent
minimum. A larger peak width indicates a wider band of particles in the vortex.
Fig.~\ref{fig:particle_Dv} shows $D_v$ changing with the particle density $\rho_0D_0^2$
and the diameter of the interaction region $d$. The ``error bars" in Fig.~\ref{fig:particle_Dv}
indicates the width of peak II, {\em i.e.}~the width of the particle band of the vortices.
In the low-density limit, the particles hardly meet each other, and the vortices mainly
contain a single SPP, which is
defined as a light-vortex-dominated state. In this state, $D_v\approx D_0$
and the width of peak II is narrow, as illustrated in Fig.~\ref{fig:particle_Dv}a for
$\rho_0D_0^2<1$. At high particle density, e.g. for $\rho_0D_0^2\geq 4$, a majority of the particles
belongs to vortices of mass larger than 10, which is defined as a heavy-vortex-dominated
state. In this state, $D_v$ still approximately equals $D_0$, but the width of peak II is
about twice as large as for $\rho_0D_0^2<1$. An extraordinarily large width at $\rho_0D_0^2=2$ in
Fig.~\ref{fig:particle_Dv}a indicates a transition of the system from a light-vortex-dominated
state to a heavy-vortices-dominated state.

The same transition from light to heavy vortices also happens with increasing $d$ at constant
density $\rho_0$.  The vortex diameters $D_v$ for different $\rho_0D_0^2$ coincide
in the heavy-vortex-dominated state, but are slightly smaller in the light-vortex-dominated state,
see Fig.~\ref{fig:particle_Dv}b. Increasing $d$ causes a systematic increase of $D_v$. The
small dip at $d=1.3D_0$, can be interpreted as follows. $D_v$ is not only determined by the
spontaneous trajectory diameter $D_0$ of circle SPPs, but also influenced by the interaction
range $d$, because the vortex prefers a dynamic temporary state with an
integral number of particle groups with a constant neighboring distance $d/2$, as shown in
Fig.~\ref{fig:particle_snapshot}. Therefore, $D_v$ can be estimated by
$D_v\approx d/[2\sin(\pi/n)]$,
where $n$ is the number of particle groups in a vortex; this estimate
provides qualitatively the correct trend for $d\ge 1.1D_0$, but deviates a little
quantitatively because $n$ differs for different vortices in the system. For example, heavy
vortices with $n=5$ are dominant when $\rho_0D_0^2=4$ and $d=1.5D_0$, while there are many
heavy vortices of both $n=5$ and $n=6$ when $\rho_0D_0^2=4$ and $d=1.3D_0$.

For $x>D_0+d$, $G_\rho$ approaches unity due to the loss of spatial correlations at large
distances, which indicates a liquid-like order of the vortices.

\begin{figure}
  \includegraphics[width=7cm]{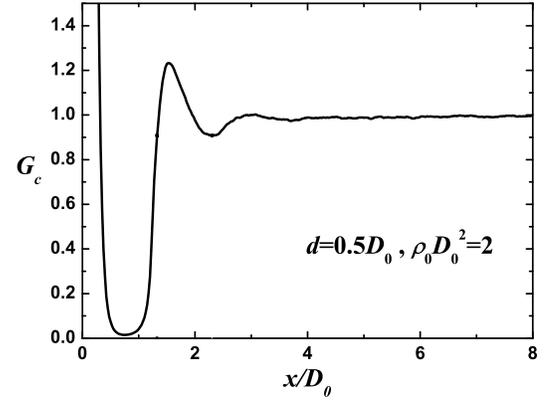}
  \caption{The trajectory-center-density correlation function, $G_c$, with
$d=0.5D_0$ and $\rho_0D_0^2=2$.}
\label{fig:particle_Gc}
\end{figure}

\begin{figure}
  \includegraphics[width=8cm]{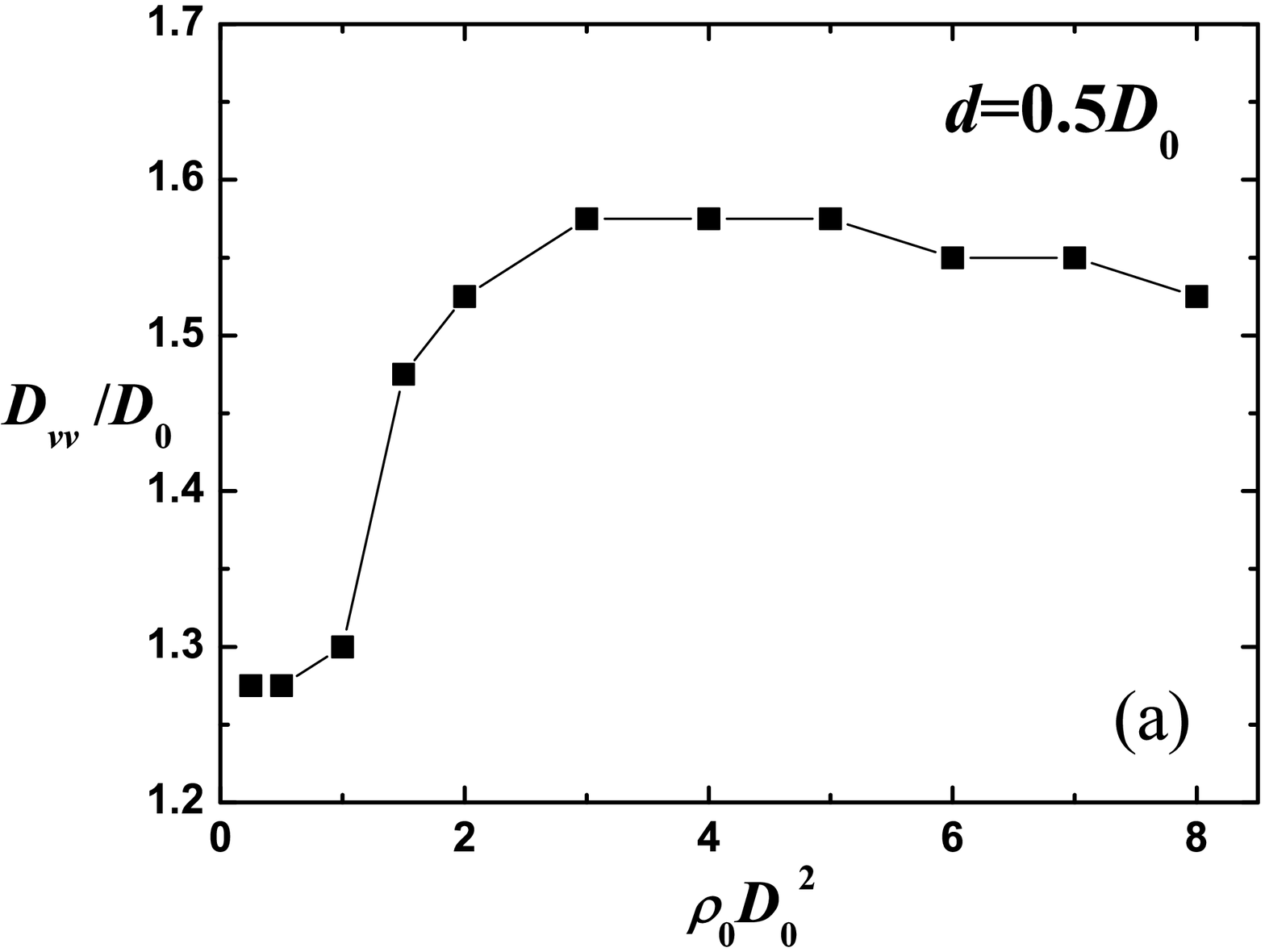}\\
  \includegraphics[width=8cm]{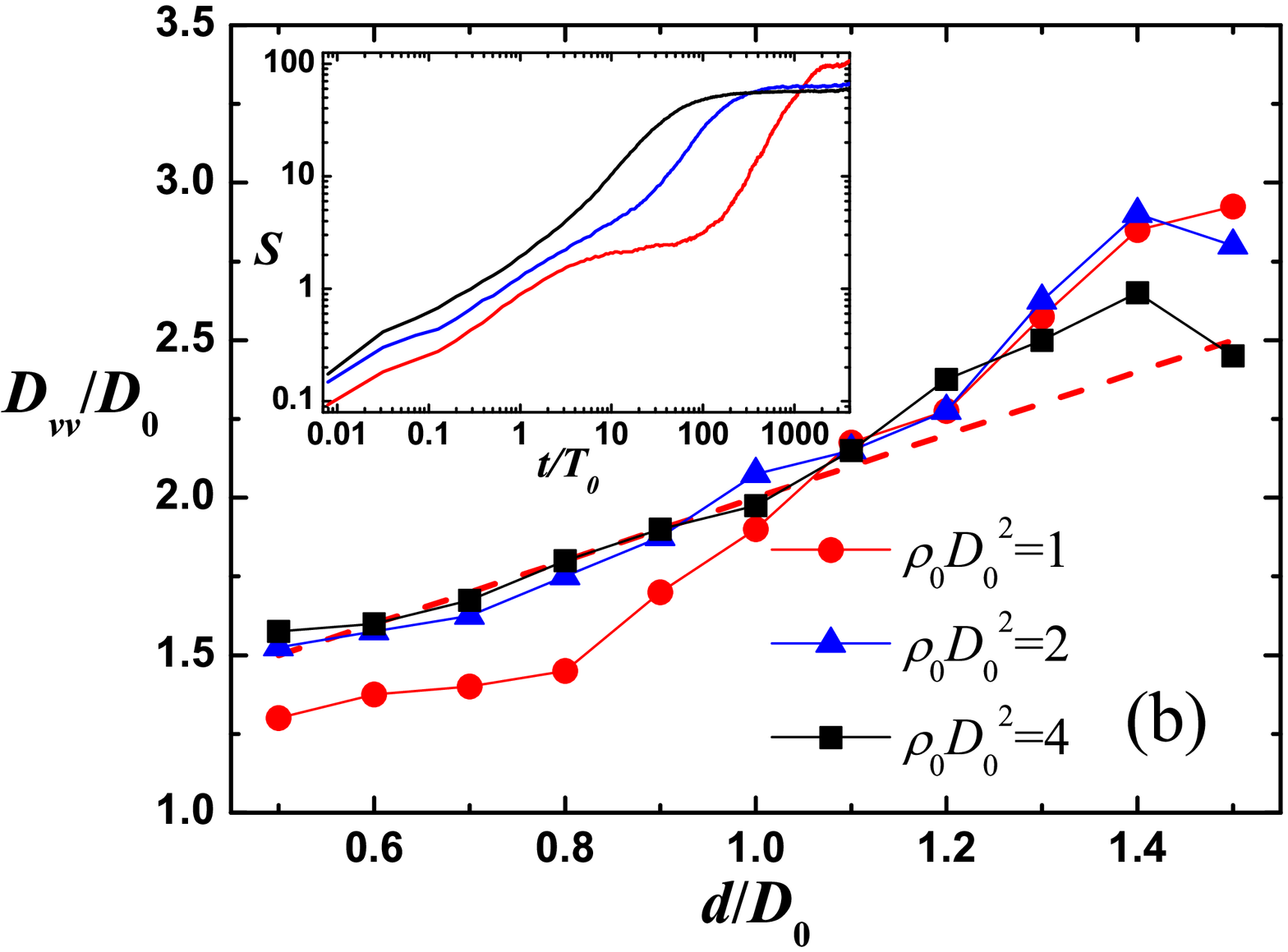}
  \caption{(Color online)(a) The distance between vortices $D_{vv}$, as a function
of the particle density $\rho_0D_0^2$. (b) $D_{vv}$ as a function of the diameter
of the interaction region $d$ with different particle densities. The red dashed line
is $y=D_0+d$. The inset is the order parameter $S$ as a function of time $t/T_0$
when $d=1.5D_0$. The red, blue and black lines
represent the particle densities $\rho_0D_0^2=1$, $2$, and $4$, respectively.}
\label{fig:particle_Dvv}
\end{figure}

The correlation function of the trajectory-center density, $G_c$, shown in
Fig.~\ref{fig:particle_Gc},
has a large peak at $x=0$ because of the aggregation of the trajectory centers. $G_c$ decays
rapidly to zero for $x \simeq 0.5D_0$, indicating that the trajectory centers aggregate in
a small spot at the center of the vortex.
Two heavy vortices cannot overlap, otherwise they will fuse into one or separate into several
distinct vortices. Therefore, $G_c$ displays a depletion zone from $D_0$ to $D_0+d/2$.
For $x \gtrsim  D_0+d/2$, $G_c$ increases rapidly and displays a peak corresponding to the
the closest possible distance $D_{vv}$ of two neighboring heavy vortices. For larger distances $x$,
$G_c$ approaches a constant with decaying oscillations, which reveals the absence of long-range
order and corresponds to a liquid-like spatial arrangement of vortices.

The dependence of $D_{vv}$ on the particle density is show in Fig.~\ref{fig:particle_Dvv}a.
$D_{vv}$ is nearly independent of density, with $D_{vv} \simeq 1.55D_0$ for $d=0.5D_0$ and
$\rho_0D_0^2\ge 3$, when the system is heavy-vortices dominated.
Similarly, $D_{vv}\simeq 1.25D_0$ for $\rho_0D_0^2\le 1$, when the system is light-vortices
dominated.
The transition between light-vortices-dominated and heavy-vortices-dominated states occurs
when $1\le \rho_0D_0^2 \le 3$, which agrees with the transition region extracted from
$G_\rho$ (compare Fig.~\ref{fig:particle_Dv}a). As shown in Fig.~\ref{fig:particle_Dvv}b,
the increase of $D_{vv}$ with $d$ for constant density $\rho_0D_0^2$ is
somewhat faster than the linear relation $D_0+d$.
For $\rho_0D_0^2=1$, $D_{vv}$ displays a pronounced increase by more than $1.5D_0$ when
$d$ is increased from $0.5D_0$ to $1.5D_0$.
When $\rho_0D_0^2=2,4$, $D_{vv}$ exhibits an unexpected decrease for $d=1.5D_0$;
we interpret this as the absence of a second vortex-mass-increasing time period
due to the inadequate simulation time, as will be explained in Sec.~\ref{sec:OP}
below.

\subsection{Order parameter}
\label{sec:OP}

In the initial state, all particles are randomly distributed in space, so that the
number of the particles and the number of the trajectory centers in a defined area
assumes a binomial distribution.  When the particles form vortices, the aggregation
of the trajectory centers cause a large deviation of the number of trajectory centers
in a defined area from its average value. A stronger aggregation leads to a larger
center-density deviation. Therefore, we define an order parameter $S$ as
\begin{equation}
S = \frac{\Delta^2}{\Delta_0^2}-1,
\label{eq:orderparameter_eq}
\end{equation}
where
\begin{equation}
\Delta_0^2=\frac{ND_0^2}{L_x L_y}\left(1-D_0^2/L_x L_y\right).
\label{eq:Delta_0}
\end{equation}
is the variance of a binomial distribution of point particles, and $\Delta^2$ is the variance of
the trajectory center numbers in an area $D_0^2$. If the center positions are randomly distributed
in the space, $S$ vanishes. $S$ increases with the degree of the aggregation of
centers, which indicates the formation of vortices. If all particles circle around a common
center, $S=N-1$.

Suppose $N$ particles form $N_v$ vortices in the two-dimensional space $L_x\times L_y$. The
number of the vortices consisting of $n$ particles is $P(n)$, and
\begin{equation}
    \sum_n P(n)=N_v,  \ \ \   \sum_n nP(n)=N.
\end{equation}
In order to elucidate the relation between the average mass $\langle n \rangle=N/N_v$
of the vortices and the order parameter, we assume that the trajectory centers of all
particles in a vortex collapse into one point, the vortex center, although they always
have some narrow distribution around the center
in the simulations. According to the trajectory-center density correlation function $G_c$
(Fig.~\ref{fig:particle_Gc}a), the distance between two vortices in our systems is usually larger
than $1.5D_0$ in the systems dominated by heavy vortices, except when two vortices are colliding
and merging. Therefore, the possibility to find two heavy vortex center in an area $D_0^2$ is low.
Thus, we assume that the probability to find two vortex centers in $D_0^2$ is zero.
We divide our system of size $L_x\times L_y$ into boxes of size $D_0^2$, and fill the boxes with
at most one vortex. Then, the variance of the center-number distribution in $D_0^2$ is
\begin{equation}
\Delta^2
=\frac{D_0^2}{L_x L_y}\left(\sum_n n^2P(n) - \rho_0 D_0^2N\right).
\end{equation}
Note that the weight average of the vortex mass is
\begin{equation}
\overline{w}=\sum_n n^2P(n)/\sum_n nP(n).
\end{equation}
Thus, $\Delta^2$ can be written in terms of $\overline{w}$ as
\begin{equation}
\Delta^2=\frac{ND_0^2}{L_x L_y}\left(\overline{w}-\rho_0 D_0^2\right).
\label{eq:Delta^2}
\end{equation}
Substituting Eqs.~(\ref{eq:Delta_0}) and (\ref{eq:Delta^2}) into Eq.~(\ref{eq:orderparameter_eq}),
we find
\begin{equation}
S=\frac{\overline{w}-\rho_0 D_0^2}{1-D_0^2/L_x L_y}-1.
\end{equation}
In our simulations, $D_0^2/L_x L_y\leq1/400\ll 1$ is negligible, so that
\begin{equation}
    S\approx \overline{w}-\rho_0 D_0^2-1.
\label{eq:S_w}
\end{equation}
This result reveals a simple linear relation between our order parameter $S$ and the weight
average of vortex mass. Thus, $S$ is a proper order parameter to characterize the degree
of vortex formation.

\begin{figure}
  \includegraphics[width=7cm]{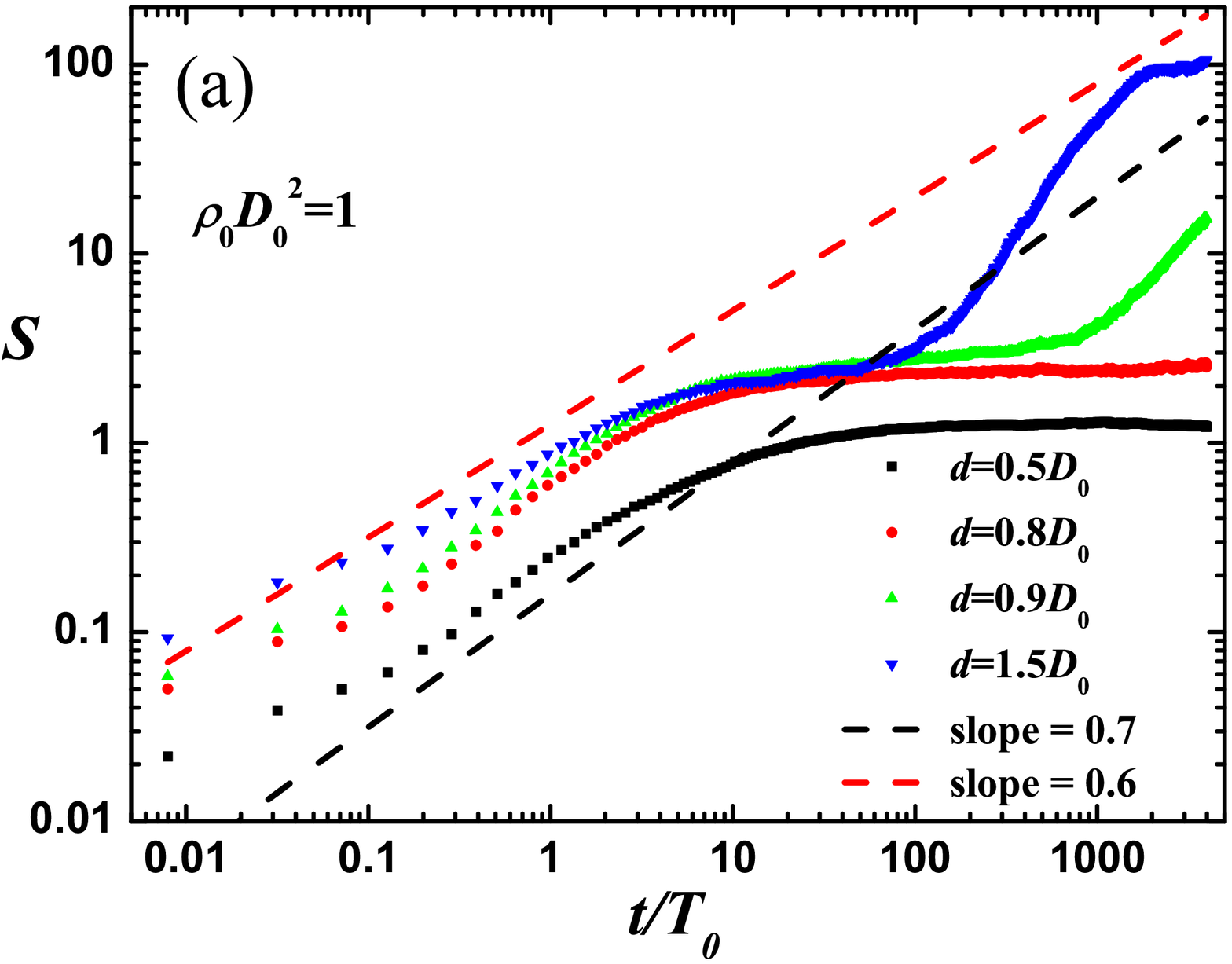}\\
  \includegraphics[width=7cm]{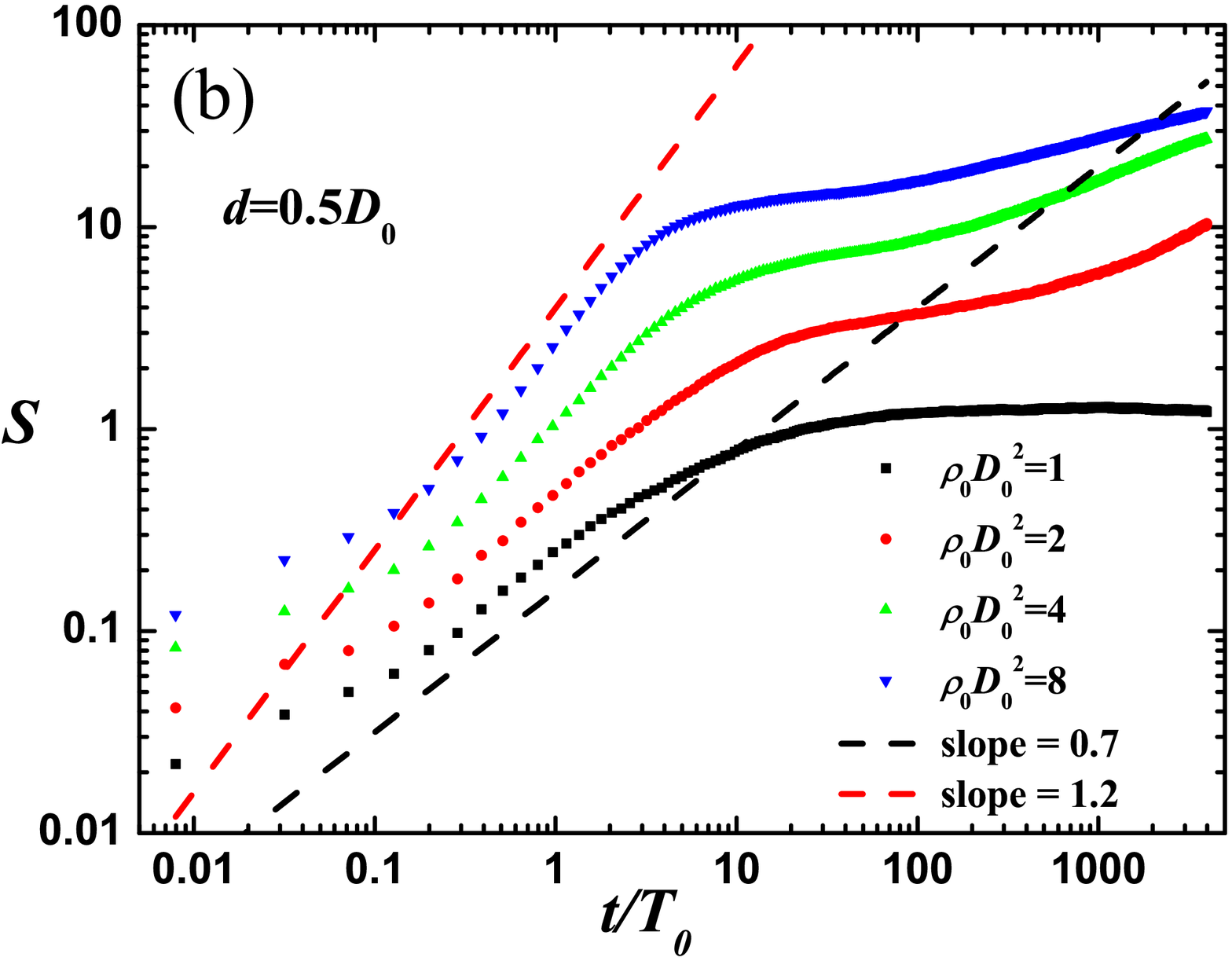}
\caption{(Color online) Time evolution of the order parameter $S$, for (a) various diameters $d$
of the interaction region, and (b) various particle densities $\rho_0$. The dashed lines are
power laws with different exponents, as indicated.}
\label{fig:particle_order}
\end{figure}

\begin{figure}
  \includegraphics[width=7cm]{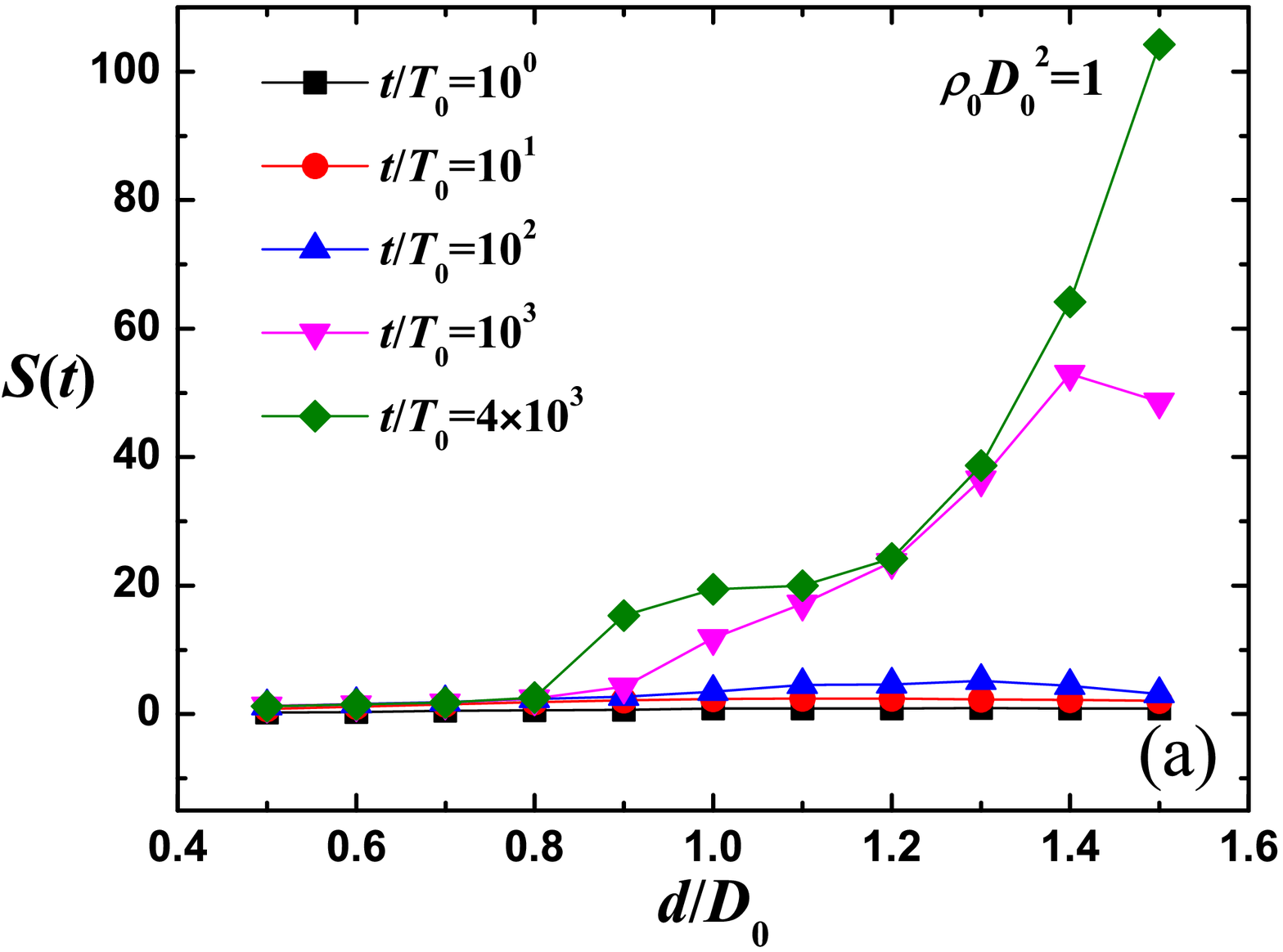}\\
  \includegraphics[width=7cm]{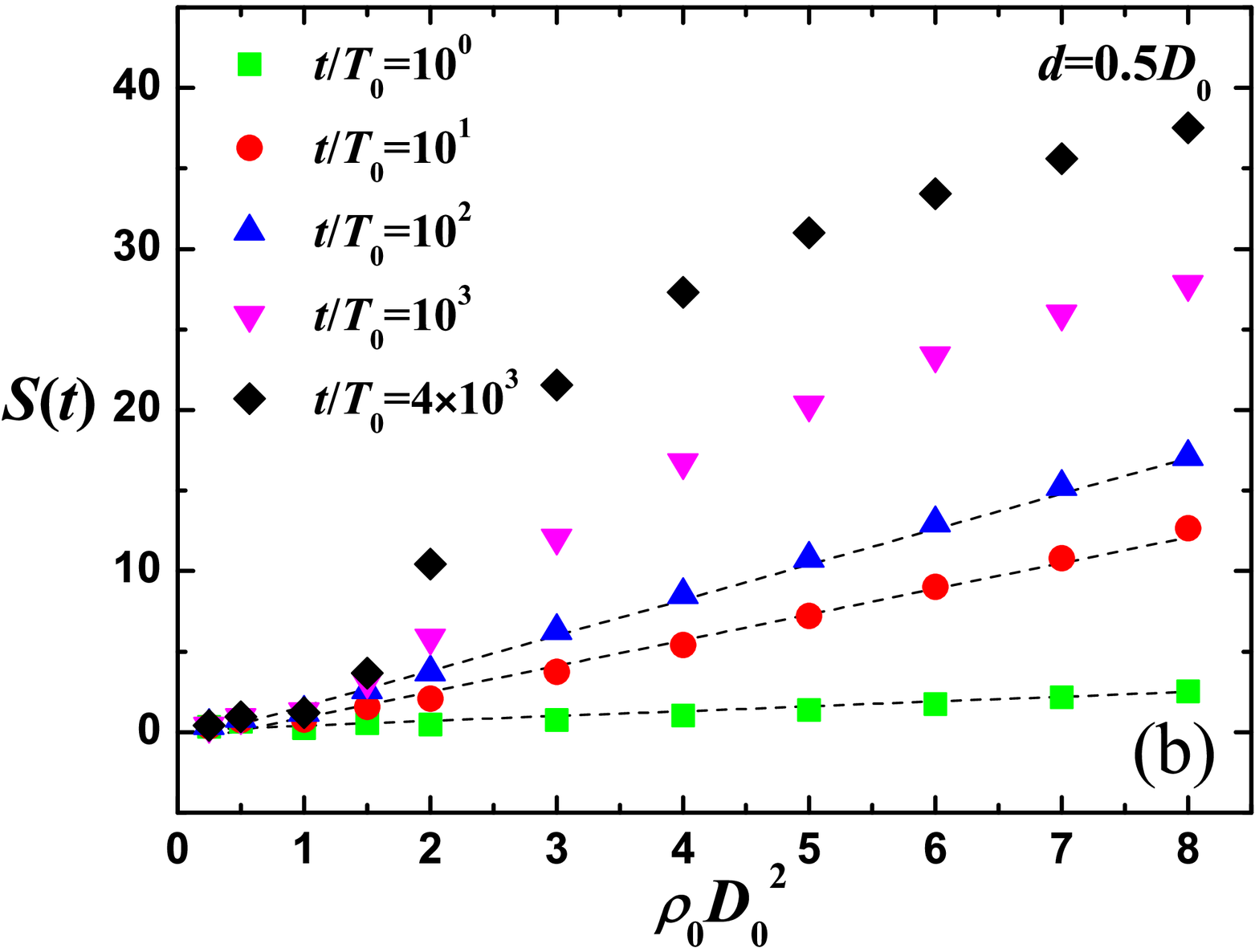}
\caption{(Color online) Order parameter $S$ at times $t/T_0=10^0$, $10^1$, $10^2$, $10^3$, and $4\times 10^3$,
(a) as a function of $d$ when $\rho_0D_0^2=1.0$, and (b) as a function of $\rho_0D_0^2$ when $d=0.5D_0$.
The dashed lines in (b) are linear fits of $S$ at $t/T_0=10^0$, $10^1$, and $10^2$.}
\label{fig:particle_order_time}
\end{figure}

The order parameter $S$ increases with time once the particle start to move and the trajectory
centers aggregate, as shown in Fig.~\ref{fig:particle_order}. $S$ first undergoes a fast increase
during a time interval which we call period I.
The increase of $S$ with time during period I occurs faster for larger $\rho_0$, but is not
sensitive to $d$, as indicated by the dashed lines in Fig.~\ref{fig:particle_order}.
During this period, the vortices are formed mainly by particles which are close to
each other in the initial state; therefore, $S$ depends roughly linearly on the
particle density, as shown in Fig.~\ref{fig:particle_order_time}b.
Period II is defined as the time period after period I, during which
$S$ increases only very slowly and approaches a plateau,
as shown in Fig.~\ref{fig:particle_order}. The plateau has a higher value for larger
$\rho_0$ and $d$.

For high densities ($\rho_0D_0^2\ge 2$ when $d=0.5D_0$ in Fig.~\ref{fig:particle_order}b)
or large $d$ ($d\ge0.9D_0$ when $\rho_0D_0^2=1$ in Fig.~\ref{fig:particle_order}a), there
is another dynamical evolution after period II, denoted period III, in which
$S$ displays another pronounced increase. The increase of $S$ in period III
indicates the formation of heavy vortices by vortex fusion --- not seen for some
low $\rho$ or small $d$ systems. However, in our simulation time scale, some systems with
high $\rho$ and large $d$ also do not display period III
(see the inset of Fig.~\ref{fig:particle_Dvv}b), in this case because the simulation time
is not long enough for vortex fusion to occur.

By analyzing the order parameter $S$ at a given time $t$ for different systems, we can also determine the
transition from the light-vortex-dominated state to the heavy-vortex-dominated state, as shown in
Fig.~\ref{fig:particle_order_time}. The transition happens when $0.8\le d/D_0\le 1.0$,
for $\rho_0D_0^2=1$, and when $1\le \rho_0D_0^2 \le 3$ for $d=0.5D_0$. $S$ assumes a small value near
zero in the light-vortex-dominated systems, then increases rapidly with increasing $d$ and $\rho_0D_0^2$
in the heavy-vortex-dominated systems.

\begin{figure}
\includegraphics[width=8.5cm]{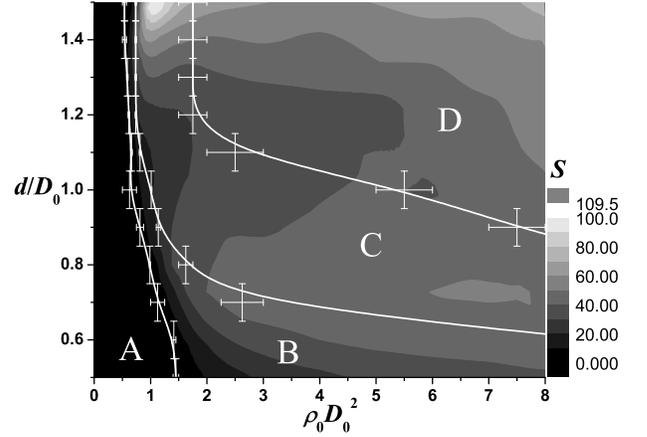}
\caption{State diagram of the circle SPPs system. The color from black to white shows the
value of $S$ at $t/T_0=4\times 10^3$ from $0$ to $10^2$. The white lines divide the state
diagram into four regimes according to the shape of $S(t)$. For the definition of regimes
$A$ to $D$ see main text.}
\label{fig:phasediagram}
\end{figure}

According to the different shapes of $S(t)$ for different systems, compare
Fig.~\ref{fig:particle_order_time}, we obtain the state diagram of
circle SPPs at $t=4\times 10^3 T_0$, as shown in Fig.~\ref{fig:phasediagram}.
The state diagram is divided into four regimes.
For systems locating in regime $A$, $S$ has completed the first growth period I, and reached the
plateau of period II, i.e.  the final state of these systems (at $t/T_0=4 \times 10^3$) is
dominated by light vortices.  In regime $B$, the systems have passed the
stationary period II, and at the end of the simulation time are just undergoing the fusion of light
vortices (period III); thus, these systems are in the crossover between regimes $A$ and $C$.
In regime $C$, the systems have come to the second stationary state after the pronounced
increase of $S$ in period III.
Finally, in regime $D$, which occupies the high $\rho$ and high $d$ section of the
state diagram,
the evolution of the fusion of heavy vortices is still in progress at $t/T_0=4 \times 10^3$;
the end state is characterized in this case by heavy vortices of similar weight.

Note that the division of the regimes according to the temporal evolution of $S(t)$
resembles but does not coincide with a classification according to the value of $S$ in the
final state state. For example, $S$ does not have the highest value in regime $D$.
The reason is that the characteristic time scale of the evolution strongly depends on particle
density, so that at the chosen final simulation time, the systems in regime $D$
has only been able to complete the initial collection of neighbor particles.
(see inset of Fig.~\ref{fig:particle_Dvv}b).

Our simulations are performed without noise. Therefore, the vortex mobility only depends on
multi-particle interactions. Increasing vortex mass leads to an increasing complexity of these
interactions and can result in a high mobility of vortices. The
light-vortex-dominated systems cannot increase the vortex mass via collision because all
vortices do not move after formation. In contrast, systems which form heavier vortices
during period I can continue to increase the vortex mass via the collision of vigorous
heavy vortices until a limit is reached, which is determined by the particle density.
At the boundary between regimes $A$ and $B$ in Fig.~\ref{fig:phasediagram}, the value of $S$
is between 1 and 3, indicating the weighted vortex mass less than 5. Moreover, the position
of the line agrees well with the transition between light-vortex-dominated state
and heavy-vortex-dominated state indicated by the variation of $D_v$ and $D_{vv}$ shown
in Fig.~\ref{fig:particle_Dv} and Fig.~\ref{fig:particle_Dvv}. Therefore, the boundary
between regimes $A$ and $B$ can be identified with the transition from the light-vortex-dominated
to the heavy-vortex-dominated state.

\section{Collective Motion of Curved, Sinusoidally-Beating Flagella}
\label{sec:flagella}

\begin{figure}
  \includegraphics[width=4cm]{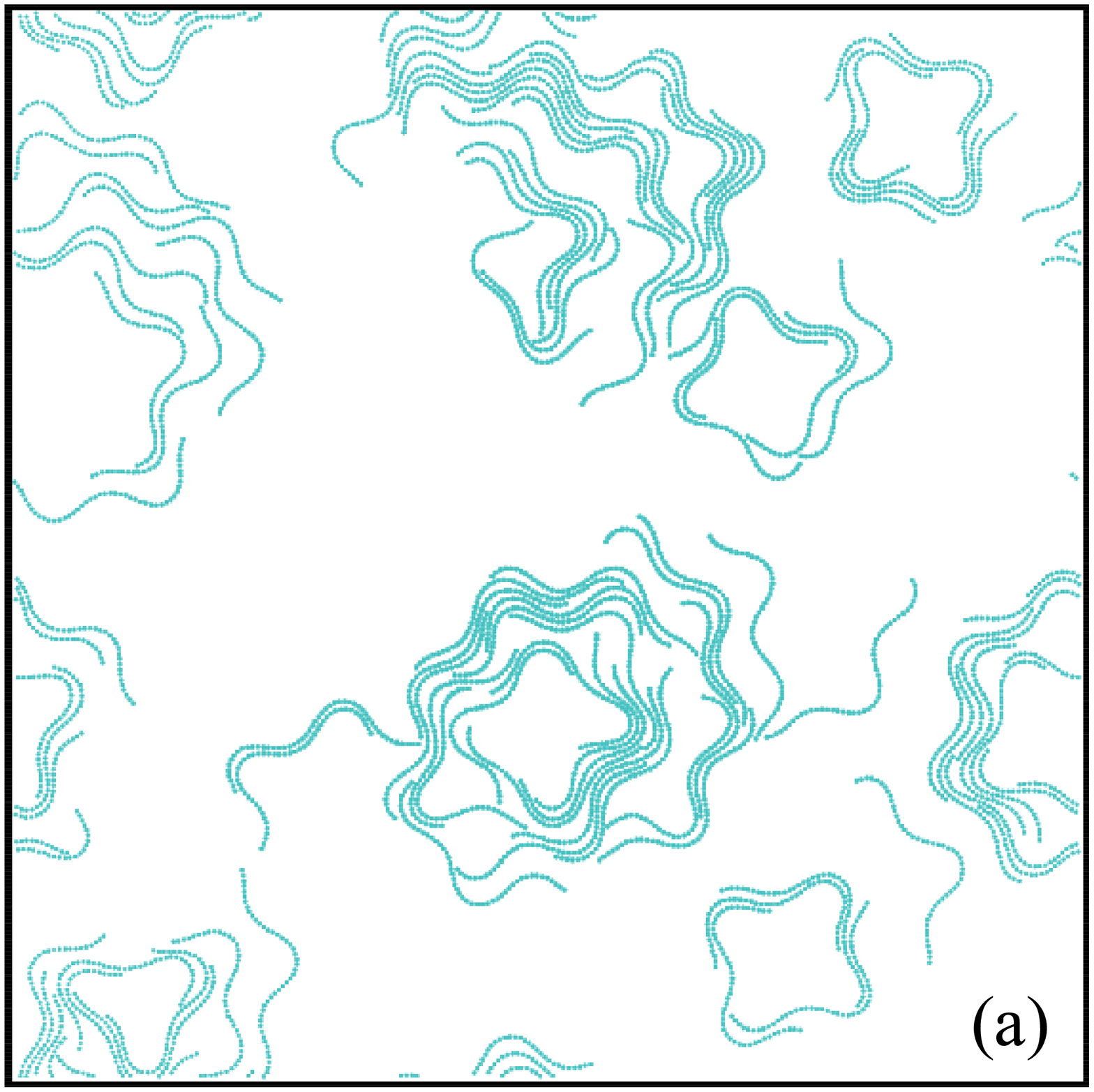}
  \includegraphics[width=4cm]{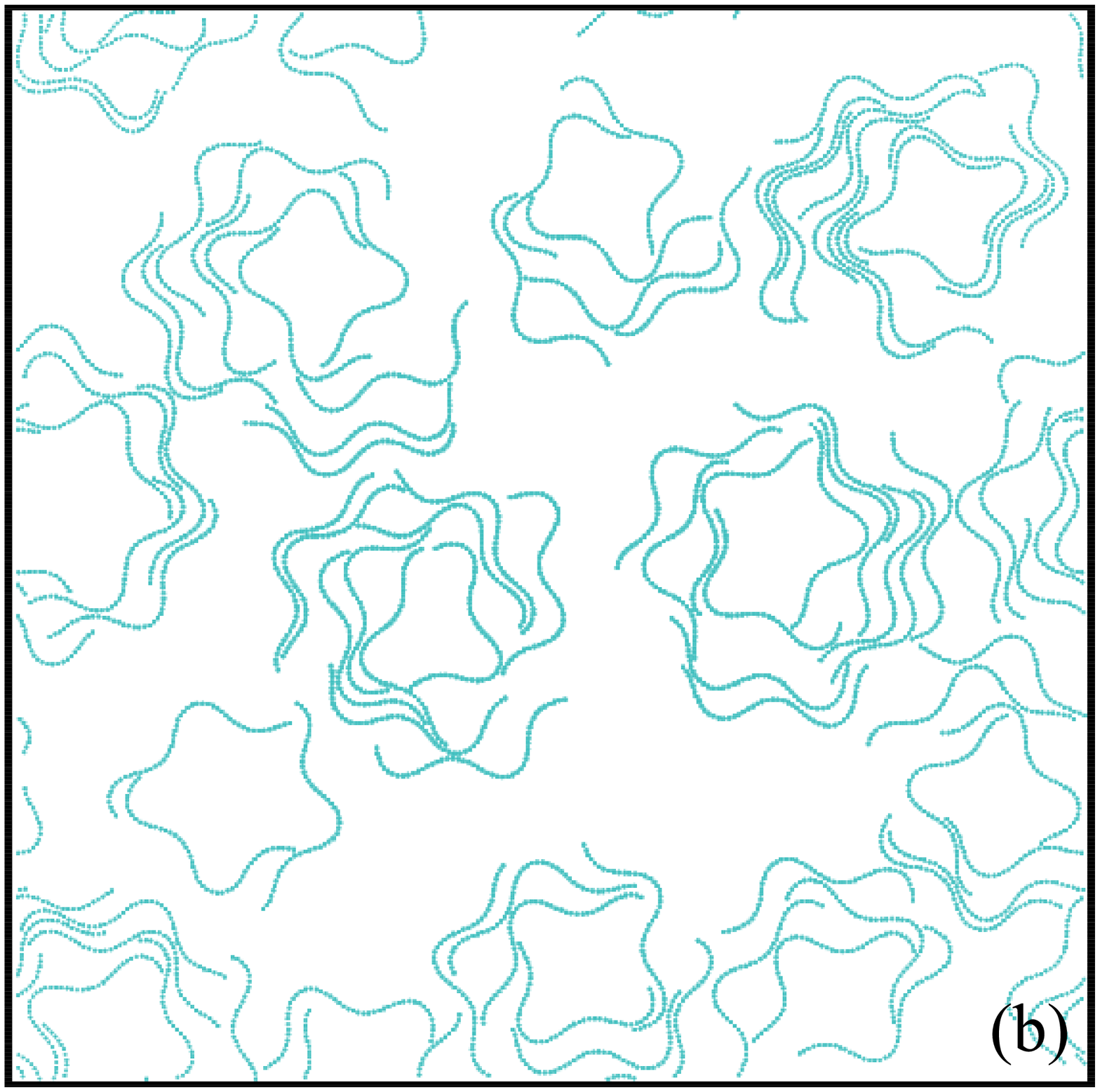}\\
  \includegraphics[width=4cm]{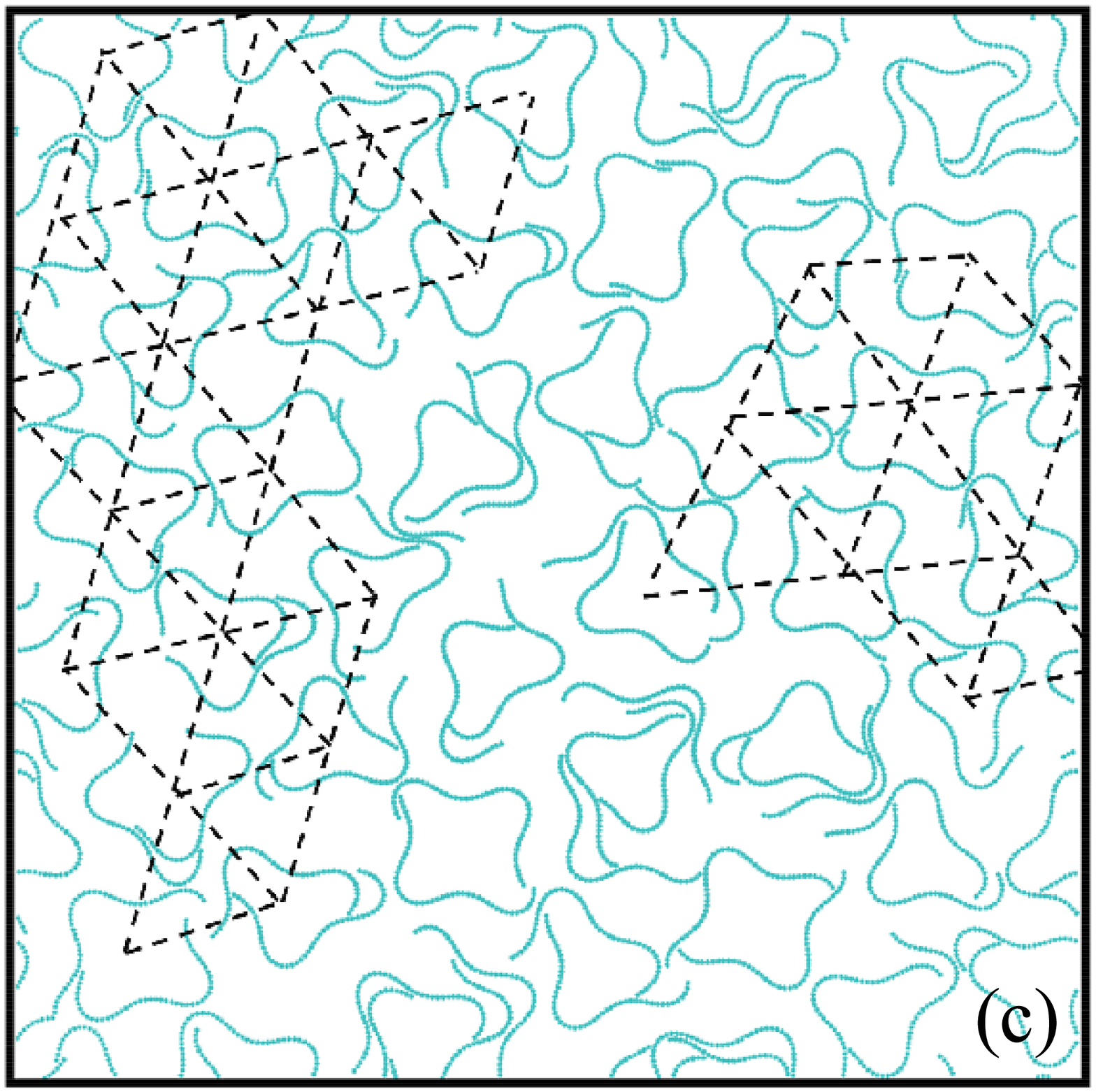}
  \includegraphics[width=4cm]{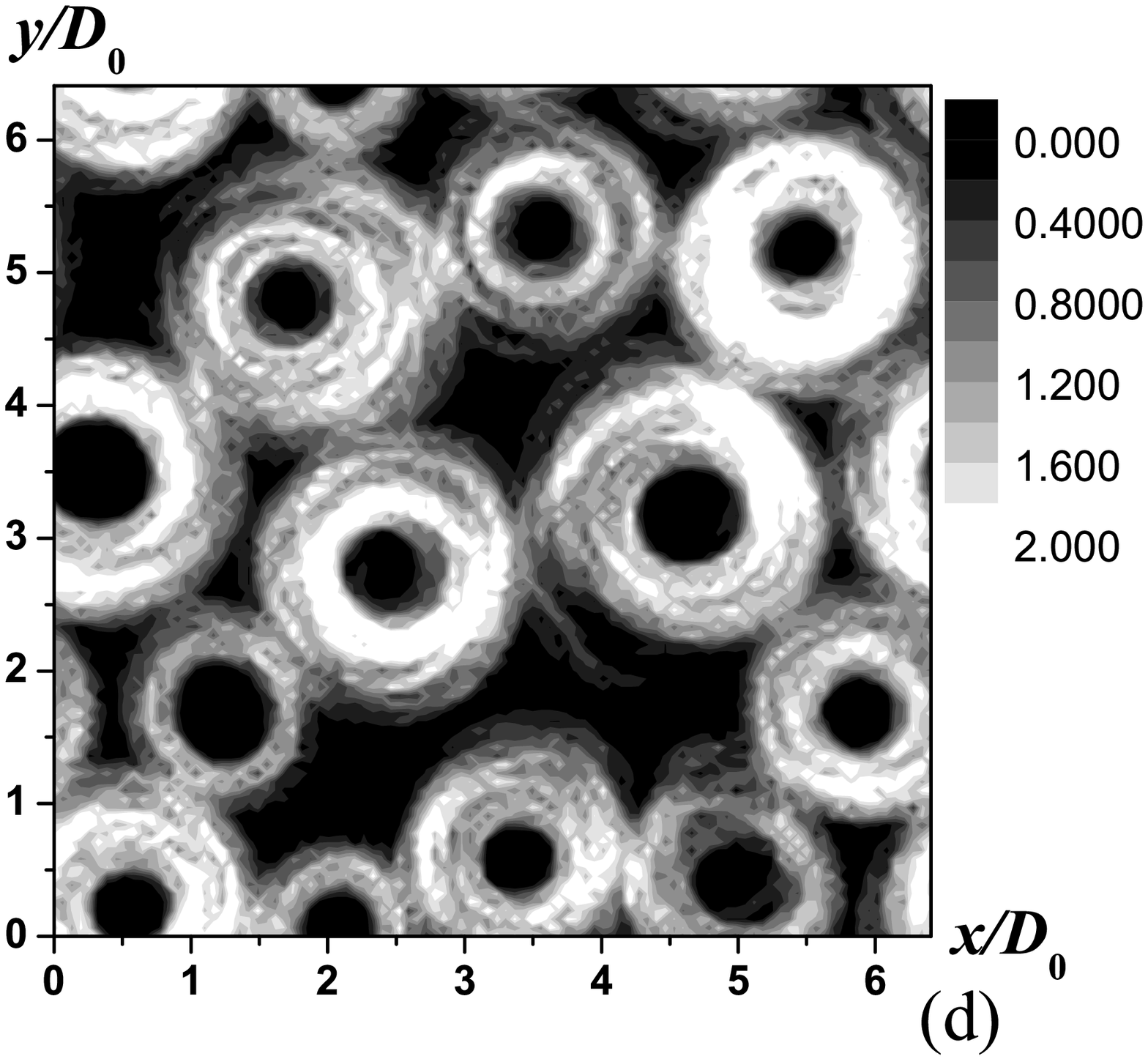}
\caption{(Color online) Snapshots of self-organized vortices of flagella (a) in a MPC fluid with
$D_0=0.614L_{fl}$, $\rho_0D_0^2=2.36$, and $\sigma=0\%$, (b) in AF with $D_0=0.614L_{fl}$, $\rho_0D_0^2=2.36$,
and $\sigma=0\%$, and (c) in AF with $D_0=0.328L_{fl}$, $\rho_0D_0^2=0.67$, and $\sigma=0\%$. The dashed
lines in (c) shows the local hexagonal order. (d) Normalized flagellum density $\overline{\rho}_f({\bf r})$
averaged over a time interval of $\Delta T=30/f_0$ of the system in (b). See also movie S1 in the
supplemental material \cite{movie}. }
\label{fig:snapshot}
\end{figure}

\begin{figure}
\includegraphics[width=7cm]{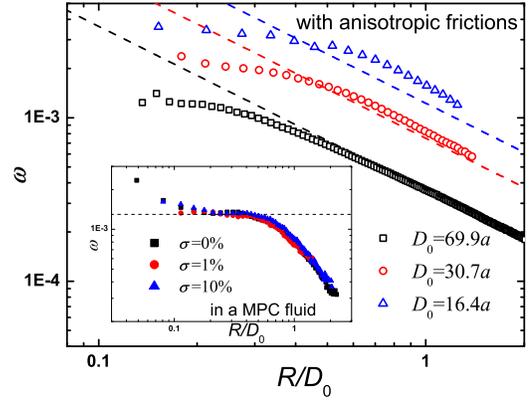}
\caption{(Color online) Angular velocity $\omega$ of the vortex as a function of
the radius position in AF simulations with $\rho_0=0.0025a^{-2}$ and $\sigma=0\%$. The dashed lines are
the functions $\omega = v_{f,0}R^{-1}$. The inset shows $\omega$ as a function of $R$ in a MPC fluid
with $\rho_0D_0^2=2.36$ and $D_0=0.614L_{fl}$; the dashed line is a guide to the eye to indicate a plateau. }
\label{fig:omega}
\end{figure}

The flagella are initially distributed randomly in space, with random orientations.
After an initial ``relaxation time", which coresponds to the time a single flagellum
needs to move several circles ($\Delta T=10\pi D_0/v_{f,0}$), the system reaches
a stationary state in which the
curved flagella spontaneously organize into rotating vortices \cite{movie}, as illustrated in
Fig.~\ref{fig:snapshot}ab. We start to gather data for averaging at $tf_0=800$, when the
flagella have completed more than ten full circles
even in the system with largest $D_0$. 
The flagella are moving clockwise and
the waves on the flagella are propagating counter-clockwise. In AF simulations, the flagella
only have hard-core interactions, while in a MPC fluid, the hydrodynamic interactions synchronize
the flagellar beat in the same vortex and packs flagella tightly due to hydrodynamic
attraction \cite{Yang2008}.
The mass of each vortex changes dynamically due to the collision with flagella in neighboring
vortices \cite{movie}.

\subsection{Angular swimming velocities}

Figure \ref{fig:omega} shows the angular velocity $\omega=v_f/R$, where $v_f$ is the center-of-mass
velocity of a flagellum and $R$ is the distance between the mass center of the flagellum and the
vortex center to which the flagellum belongs, as a function of $R$. In AF simulations, $\omega$
approaches $v_{f,0}/R$ for large $R$, where $v_{f,0}$ is the velocity of a freely swimming
flagellum. At small radii, the volume exclusion between the propagating sinusoidal
configurations of neighboring flagella reduces the angular velocity. Thus, although $\omega$
still slowly increases with decreasing $R$ for $R<0.5D_0$, it is much lower than expected
from the relation $\omega=v_{f,0}/R$. On the other hand, at large radius ($R>0.5D_0$),
in some AF systems (for example, $D_0=0.614L_{fl}$ and $D_0=0.328L_{fl}$ in Fig.~\ref{fig:omega}),
$\omega$ is larger than $v_{f,0}/R$ due to the repulsive interaction between two flagella
belonging to different vortices.  In conclusion, the volume exclusion between the flagella
depresses $\omega$ due to interactions between flagella in the same vortex, but
enhances $\omega$ due to interactions between neighboring vortices.

In a MPC fluid, $\omega$ also approximately obeys the power law $\omega \sim R^{-1}$ at
large $R$. However, at radii $0.2<R/D_0<0.5$, $\omega$ is nearly independent of $R$, as shown in
Fig.~\ref{fig:omega}b. Here, the synchronized flagella form a closed ring with integer
numbers of waves and rotate with same angular velocity, as shown in movie S2 in the
supplementary material \cite{movie}.

Although $\omega R$ is not exactly equal to $v_{f,0}$ in systems with MPC fluid or with AF,
an assumption of a unique swimming velocity is still a good approximation for comparison
of the flagella system with the circle SPPs system.

\subsection{Correlation functions}

We define a normalized density of flagellar segments as
\begin{equation}
\overline{\rho}_f({\bf r},t)=\frac{1}{\Delta T}\int_{t-\Delta T/2}^{t+\Delta T/2}{\bf d}t \,
                                        \frac{\rho_f({\bf r},t)}{\rho_{f,0}},
\end{equation}
where $\rho_f({\bf r},t)$ is the number density of monomer beads
averaged in a square box of area $(L_x/100) \times (L_y/100)$ at the position ${\bf r}$ at time $t$.
In order to gain better statistics of the vortex structures, $\overline{\rho}_f$ is the
average over time $\Delta T$, with $\Delta T$ is chosen to be $30/f_0$. An example of an image of
$\overline{\rho}_f({\bf r},t)$ is shown in Fig.~\ref{fig:snapshot}d.

\begin{figure}
  \includegraphics[width=7cm]{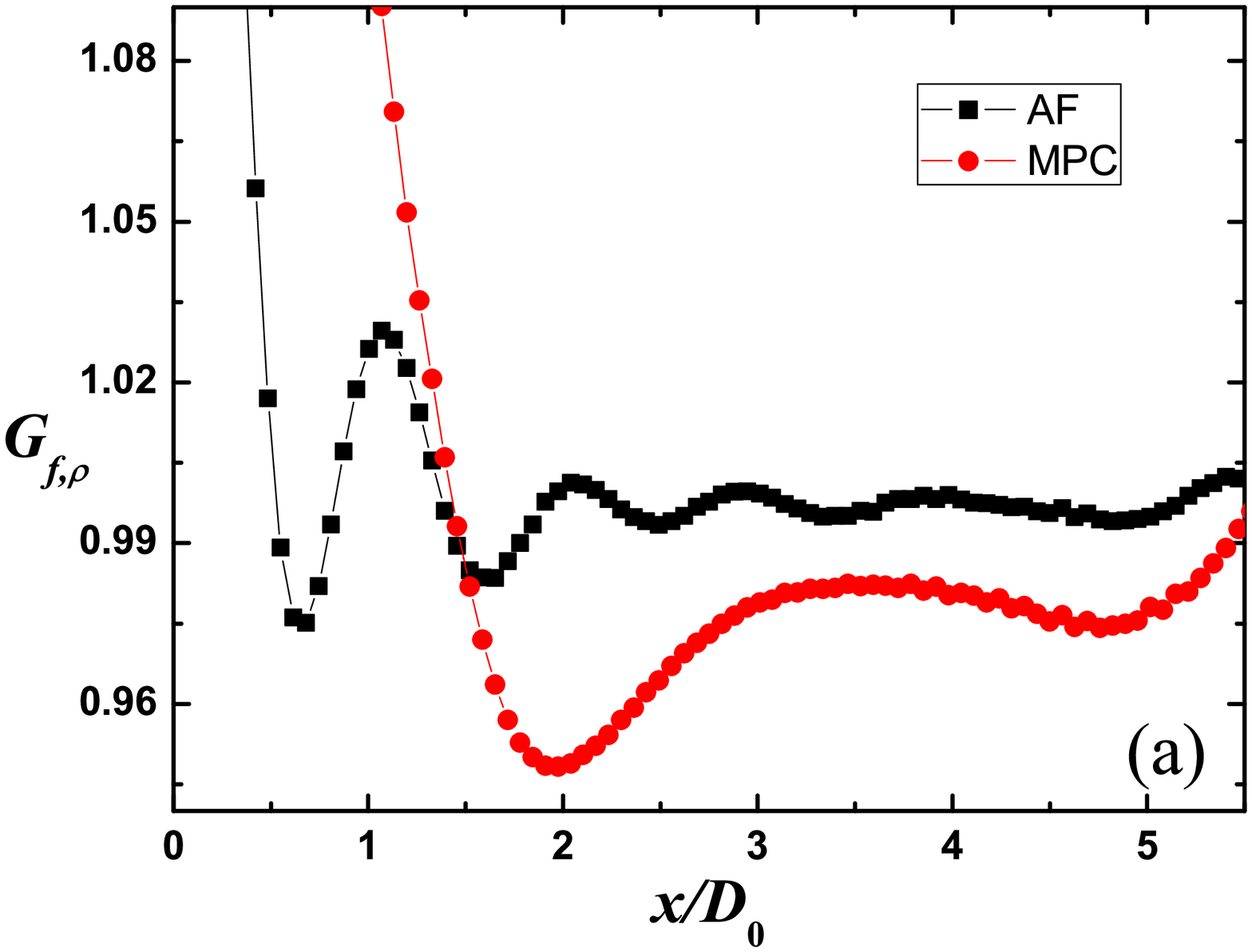}\\
  \includegraphics[width=7cm]{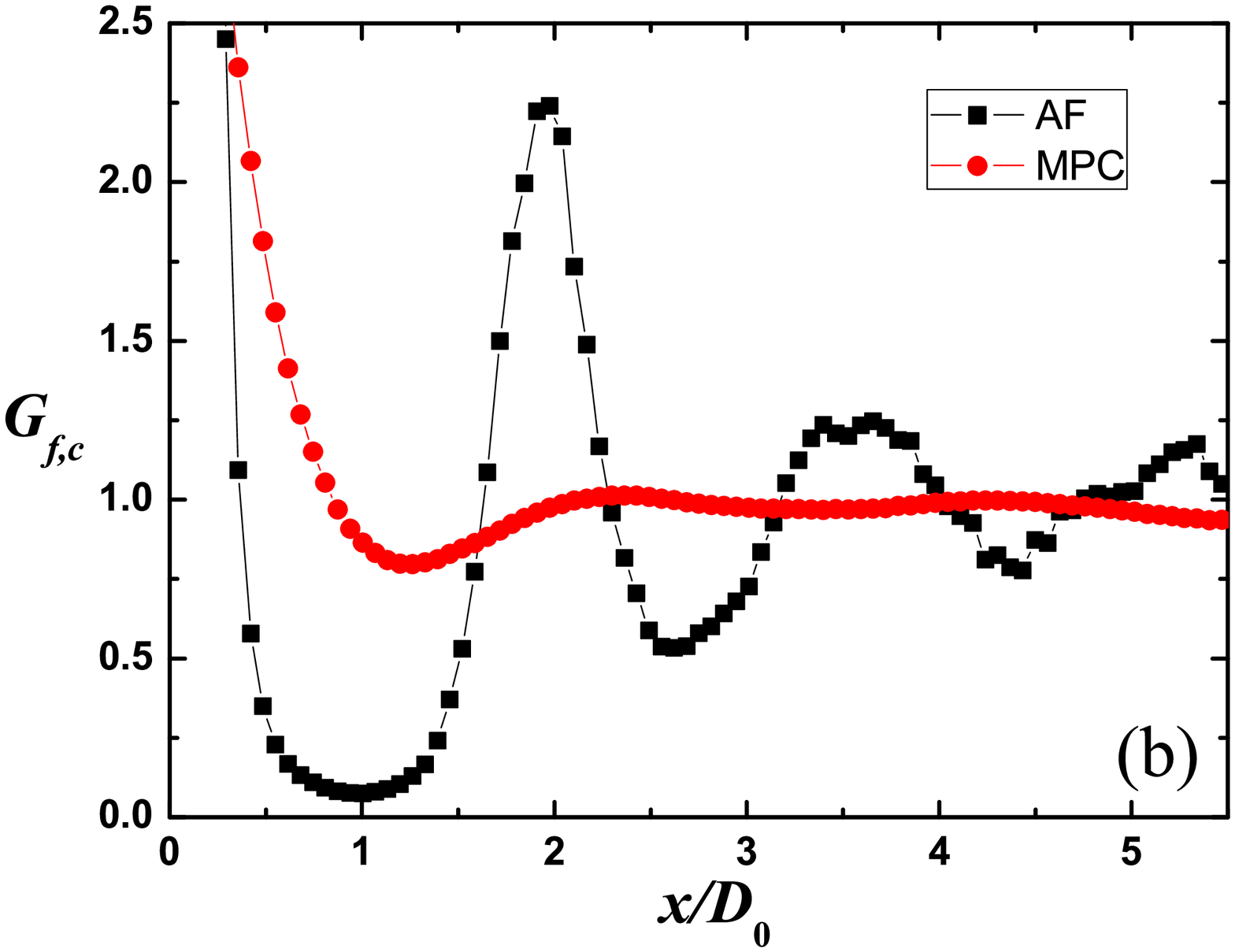}
\caption{(Color online) Correlation functions of (a) the normalized flagellum density
$\overline{\rho}_f({\bf r})$ and (b) the normalized trajectory center density
$\overline{\rho}_{f,c}({\bf r})$.
Both densities $\overline{\rho}_f({\bf r})$ and $\overline{\rho}_{f,c}({\bf r})$ are averaged
over a time interval $\Delta T=30/f_0$. The parameters are $D_0=0.614L_{fl}$, $\rho_0D_0^2=2.36$,
and $\sigma=0\%$. The symbols indicate the results of MPC-fluid (red bullets) and of
AF (black squares) simulations. }
\label{fig:correlation}
\end{figure}

The correlation function of flagellum density is then defined as
\begin{equation}
G_{f,\rho}(|{\bf r}-{\bf r}'|)=\langle\overline{\rho}_f({\bf
r},t)\cdot\overline{\rho}_f({\bf r}',t)\rangle_t.
\end{equation}
Similarly, the correlation function of flagellum trajectory-center density is
\begin{equation}
G_{f,c}(|{\bf r}-{\bf r}'|)=\langle\overline{\rho}_{f,c}({\bf
r},t)\cdot\overline{\rho}_{f,c}({\bf r}',t)\rangle_t.
\end{equation}
where $\bar\rho_{f,c}({\bf r},t)=\rho_{f,c}({\bf r},t)/\rho_{f,0}$ and
$\rho_{f,c}({\bf r})$ is the number density of the centers of flagellum trajectories
at ${\bf r}$ and time $t$.
Figures~\ref{fig:correlation}a,b show examples of $G_{f,\rho}$ and $G_{f,c}$, respectively.
Both correlation functions approach a constant at large distances, indicating
the absence of long-range order and a liquid-like arrangement of vortices,
similarly as discussed for circle SPPs in Sec.~\ref{sec:SPP_corr}.
Long-range order of vortices, such as a hexagonal arrangement, is not observed
even for large flagellum densities. However, a local hexagonal order is still possible
due to the volume exclusion between vortices, as indicated by the higher-order peaks of
$G(r)$ and shown in real-space snapshots in Fig.~\ref{fig:snapshot}c.

The interpretation of the correlation function is of course very similar as for
correlations functions of circle SPPs in Sec.~\ref{sec:SPP_corr}. The first local
maximum of $G_{f,\rho}$ corresponds to the average vortex diameter, the first local
maximum of $G_{f,c}$ to the average distance between neighbor vortices.
A comparison of AF and MPC simulations clearly shows that structures in the vortices
in the AF model are considerably more ordered, which leads to pronounced oscillations
of the correlations functions.  In the AF simulations, $G_{f,c}$ nearly vanishes
at $x\simeq D_0$, indicating that the area occupied by the trajectory centers in a
vortex is significantly smaller than $D_0$, in agreement with our
circle SPPs observations. In the MPC-fluid simulations, the weaker correlations indicate
a larger diversity of vortex sizes but also a larger mobility of the vortices.

\begin{figure*}
  \includegraphics[width=5.5cm]{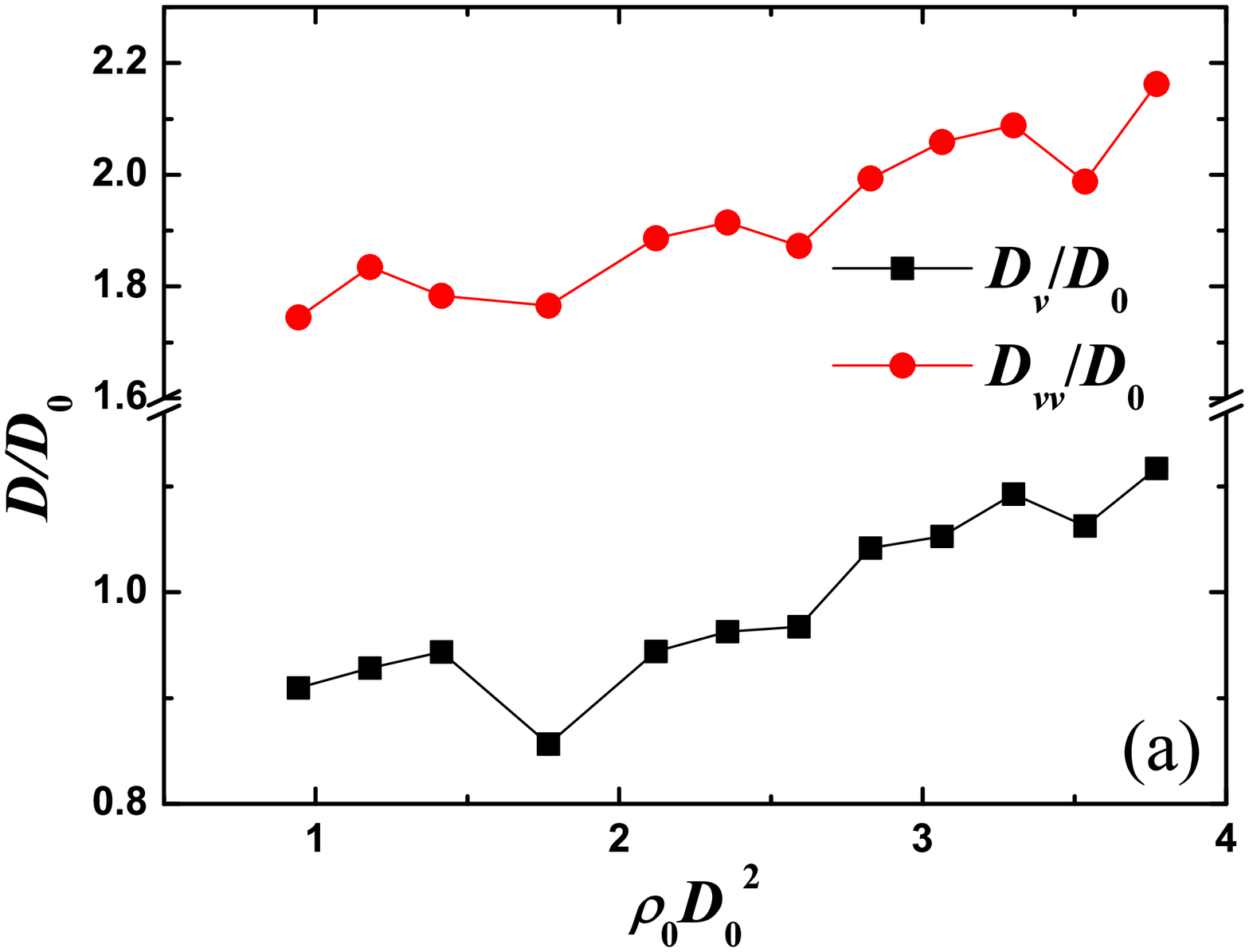}
  \includegraphics[width=5.5cm]{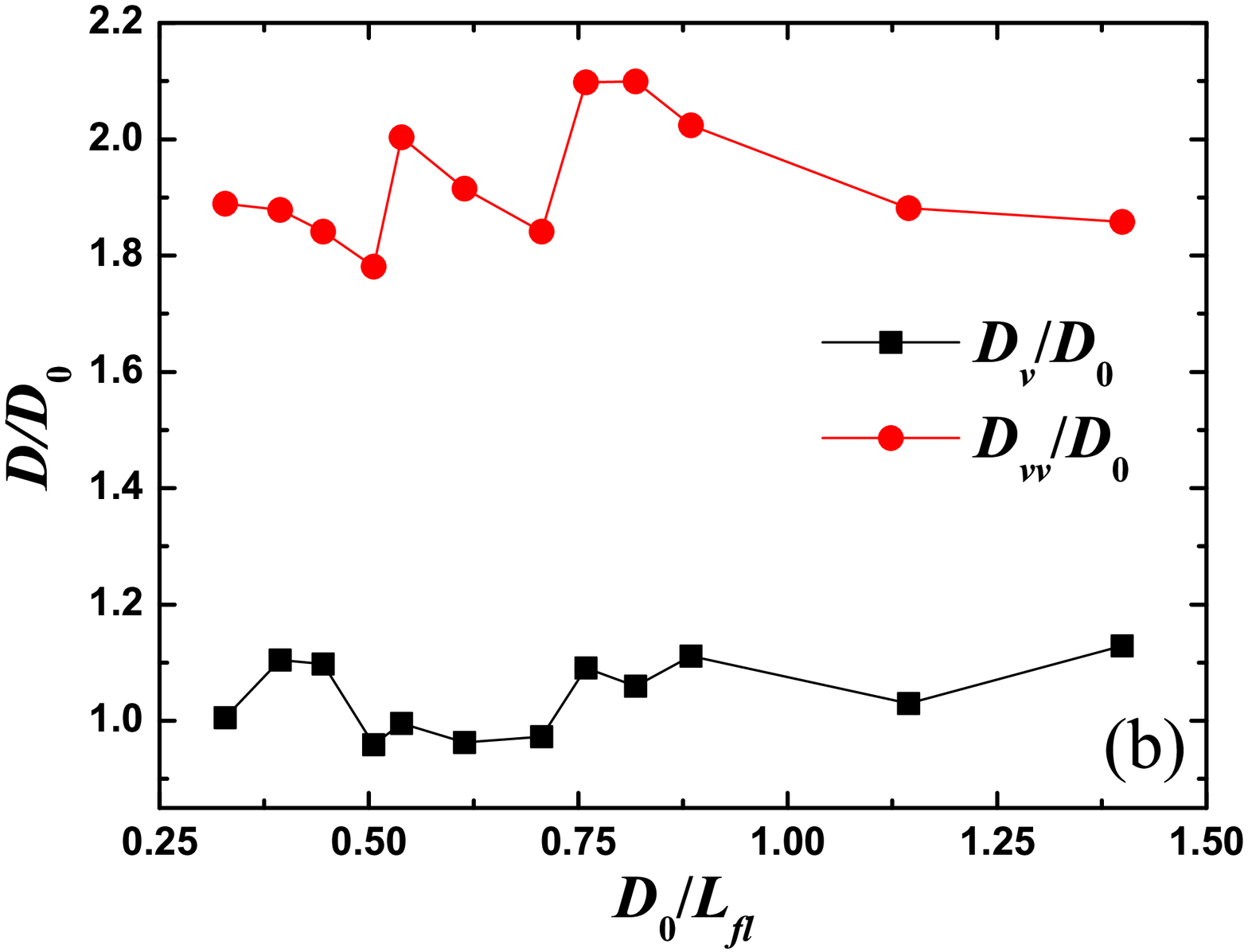}
  \includegraphics[width=5.5cm]{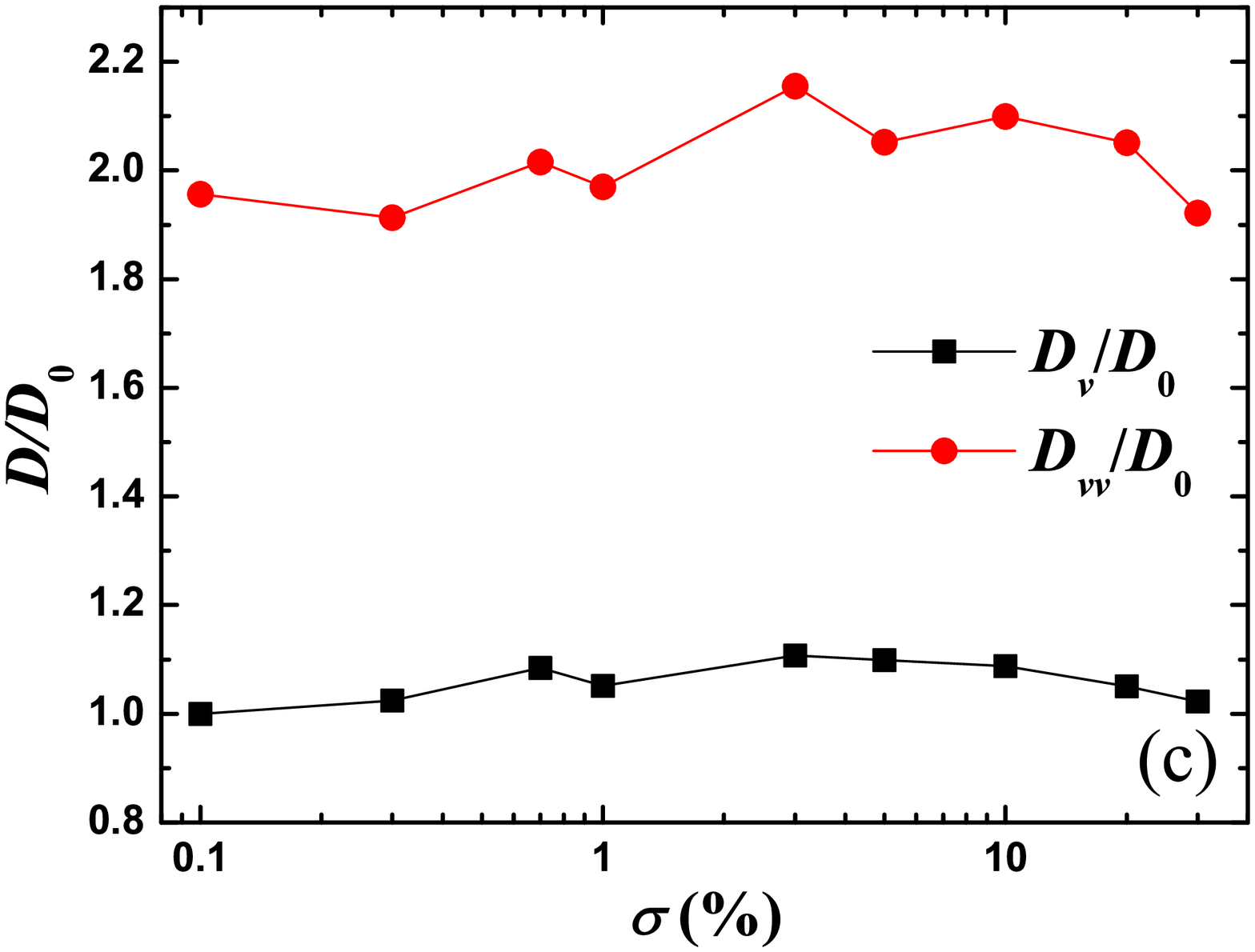}
\caption{(Color online) Vortex diameter $D_v$ and average distance between vortices
$D_{vv}$ as a function of (a) the flagellum density $\rho_0$, (b) the spontaneous
trajectory diameter $D_0$, and (c) the variance $\sigma$ of frequency distribution
in AF simulations.}
\label{fig:D1D2}
\end{figure*}

A comparison of the snapshots and correlation functions in Figs.~\ref{fig:snapshot}
and \ref{fig:correlation}b with those of sea-urchin sperm vortices in
Ref.~\onlinecite{Riedel2005} shows that the phenomena observed in AF simulations
are more similar to the experimental behavior than those in MPC simulations.
Furthermore, the results of our AF simulations agree much better with those of
our circle SPPs simulations. We attribute the deviations of the MPC-simulation results
from those of the experiments to the very strong hydrodynamic interactions in the
two-dimensional model system. Indeed, the flow field around a dragged point particle in two
spatial dimensions
decays only logarithmically with distance $r$, while in three dimensions it decays
much faster, like $1/r$, and even faster in the presence of a wall.  In the experiments,
the motion of sperm near a wall is governed by three-dimensional hydrodynamics. Thus,
we conclude that a detailed numerical investigation of the importance of hydrodynamic
interactions for the formation of sperm vortices
requires full three-dimensional hydrodynamic simulations. The two-dimensional MPC simulation
still provide the important information that the synchronized flagellum
beating of the sea-urchin sperms in a vortex is the result of hydrodynamic interactions
between the beating tails, as indicated by the snapshot in Fig.~\ref{fig:snapshot}a
(see also Refs.~\onlinecite{Yang2008,Yang2010}).
In the remainder of this section, we focus on the analysis of the systems
with anisotropic friction.

\subsection{Vortices of flagella in the anisotropic-friction model}

Figure~\ref{fig:D1D2} shows the average vortex diameter $D_v$ and the average vortex
distance $D_{vv}$ as functions of $\rho_0$, $D_0$, and $\sigma$. In the low-density
limit, the flagella move without touching each other. Thus $D_v$ must approach
$D_0$ and $D_{vv}$ must approach the average distance between flagella.
As the density increases, $D_v$ and $D_{vv}$ increase slowly, as shown in
Fig.~\ref{fig:D1D2}a. As in the circle SPPs systems,
compare Figs.~\ref{fig:Graph_particle_Grho} and \ref{fig:particle_Dvv},
$D_v$ and $D_{vv}$ also increase and level off when $2\le \rho_0D_0^2 \leq 4$.
However, the drop of $D_v$ and $D_{vv}$ near $\rho_0D_0^2=2$ for circle SPPs systems,
which indicates the transition from the light-vortex-dominated to the
heavy-vortex-dominated state, is not seen in the curved-flagella system.
On the other hand, the increase of $D_v$ can also partially be attributed to
the volume exclusion between flagella, so that the orbit at radius $R$ in
the vortex can be occupied only by a limited number of flagella. Similarly, $D_{vv}$
increases with $\rho_0D_0^2$ due to volume exclusion, which generates an
effective repulsion between the neighboring vortices.

When we change the preferred trajectory diameter $D_0$ of the curved flagella,
the diameter of a vortex varies as $D_v\approx D_0$, as
shown in Fig.~\ref{fig:D1D2}b, in good agreement with the behavior of the
circle SPPs systems. However, over a wide range of $\rho_0$ or $D_0$,
$D_{vv}/D_0$ remains nearly independent of these parameters and fluctuates
around a value $1.9\pm0.2$. This value of $D_{vv}/D_0$  is found for
the interaction range $d/D_0=0.9$ to $1.0$ in circle SPPs systems
(Fig.~\ref{fig:particle_Dvv}b).
Thus, we conclude that the effective size of the interaction region of a curved
flagellum is approximately the same as the diameter of its circular trajectory.

In the sea-urchin sperm vortex
experiment \cite{Riedel2005}, the radius of the sperm vortex is
$D_v/2=13.2 \pm 2.8 \mu$m, and the average vortex distance is $D_{vv}=49 \pm 9 \mu$m.
Therefore, $D_{vv}/D_v=1.86 \pm 0.52$ in the experiment, in excellent agreement with
$D_{vv}/D_v=1.8\pm0.2$ in our AF simulations. In our circle SPPs simulations,
$D_{vv}/D_v=1.83\pm0.10$ for systems with $d=0.9D_0$ in the heavy-vortex-dominated state.

The vortex formation is not sensitive to the width of the beating-frequency
distribution of the flagella, as shown in Fig.~\ref{fig:D1D2}c, although it
leads to a range of flagellar velocities. This
insensitivity explains the emergency of the vortices in the sea-urchin sperm
experiment even though there is a spread of beating frequency of about 9 percent
\cite{Riedel2005}.

\begin{figure*}
  \includegraphics[width=5.5cm]{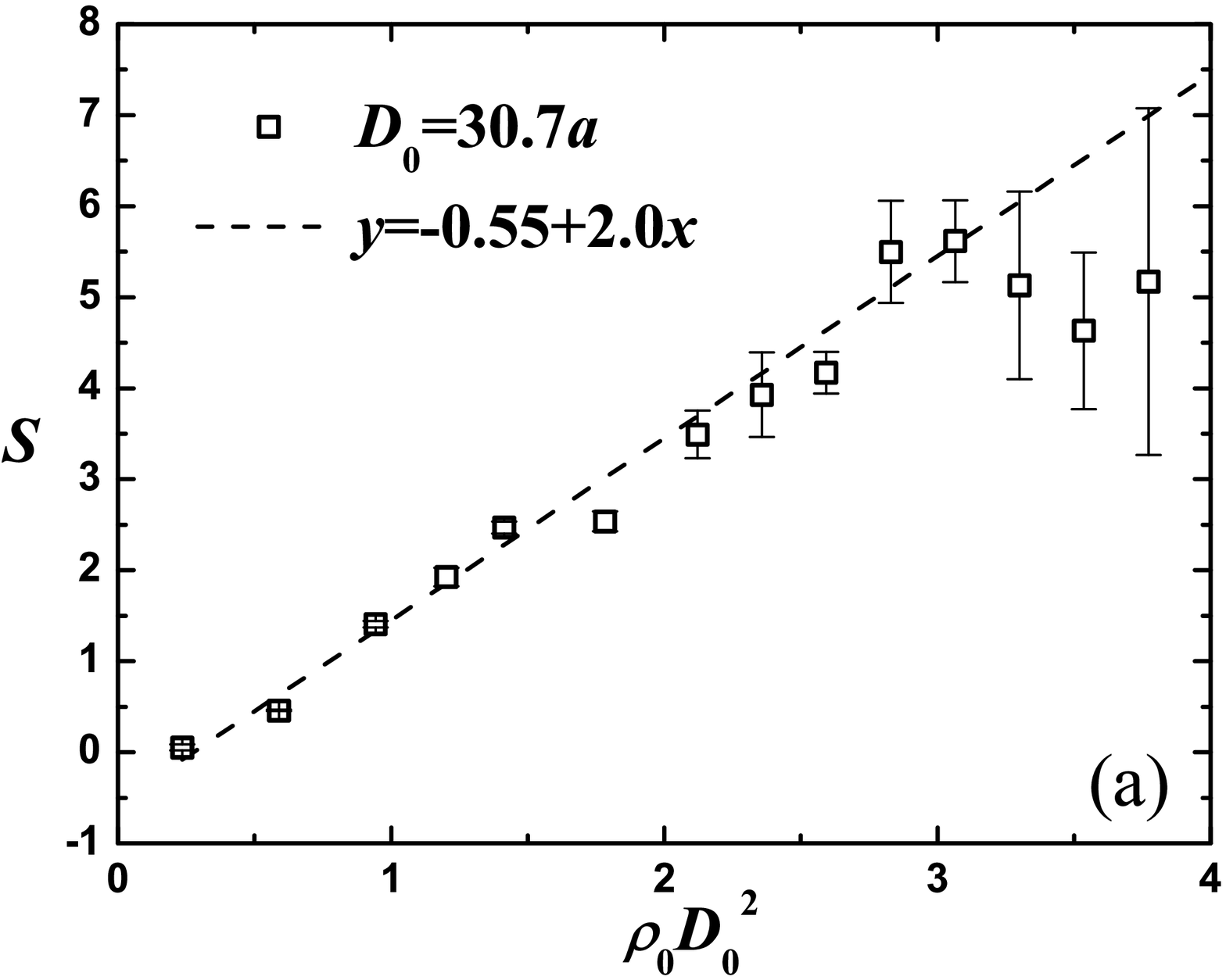}
  \includegraphics[width=5.5cm]{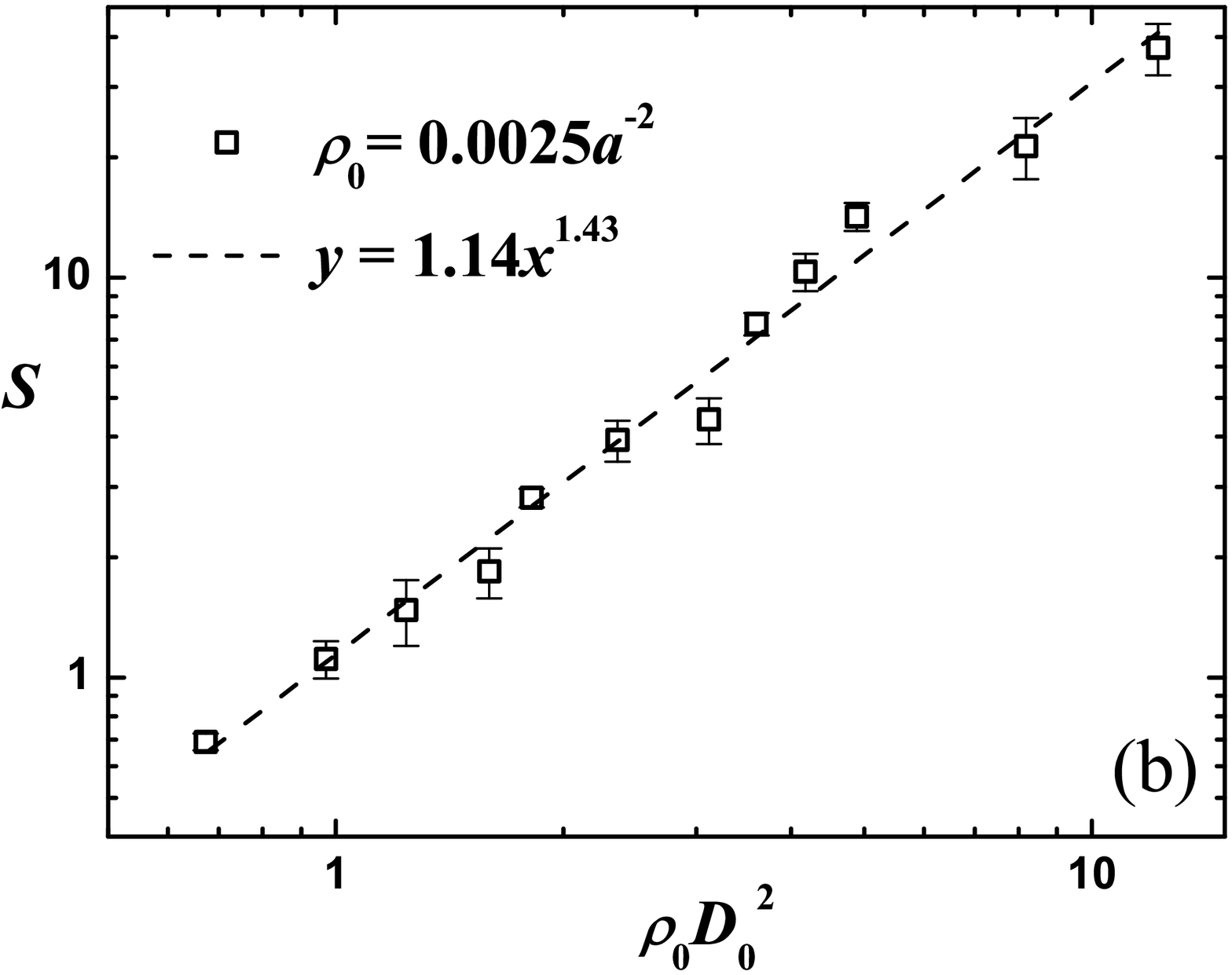}
  \includegraphics[width=5.5cm]{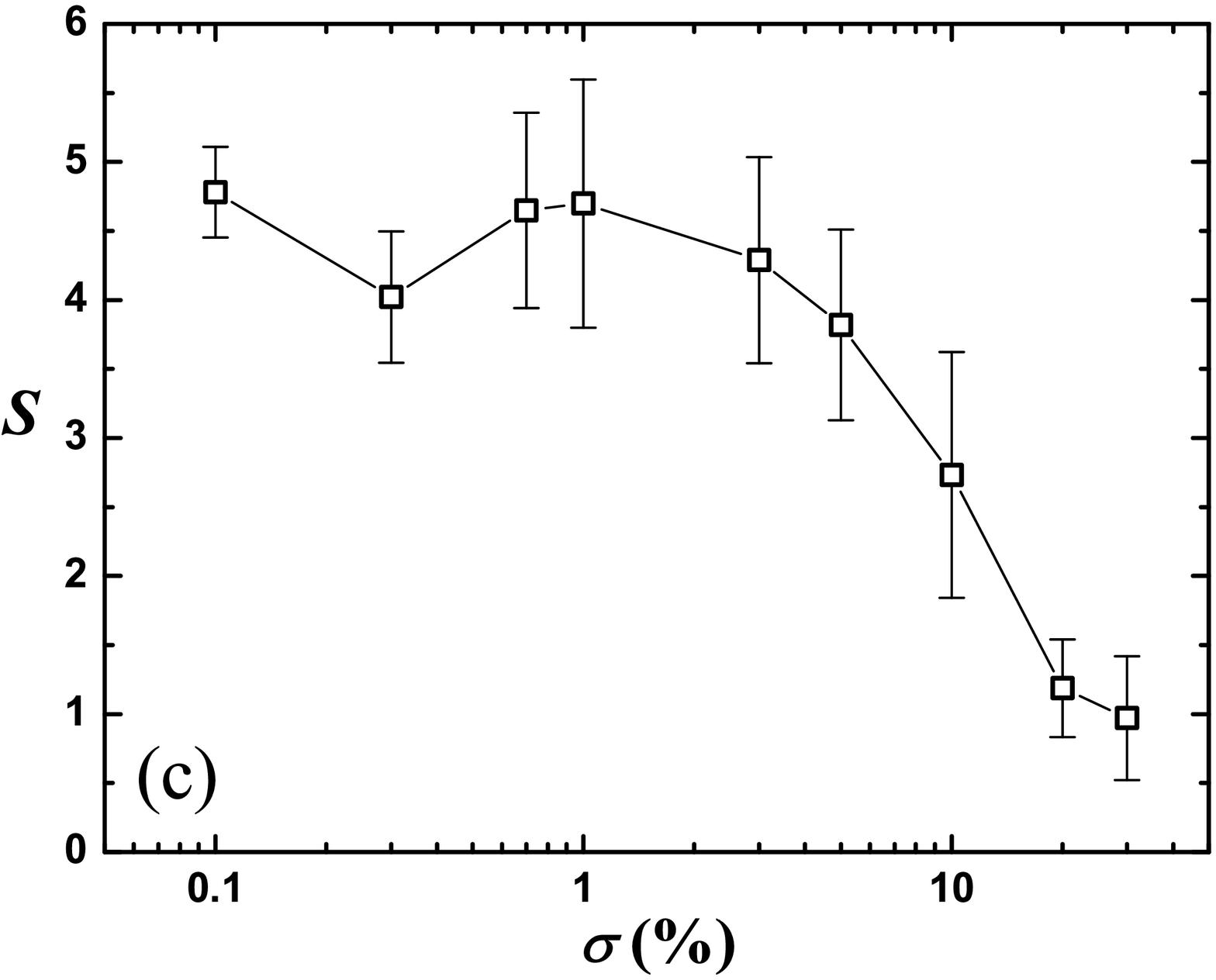}
\caption{Order parameter $S$ as a function of (a) the scaled density $\rho_0 D_0^2$,
with $D_0=0.614L_{fl}$ and $\sigma=0\%$,
(b) the scaled circle area $\rho D_0^2$, with $\rho_0=0.0025a^{-2}$ and
$\sigma=0\%$, and (c) the width $\sigma$ of the freqency distribution,
with $\rho_0D_0^2=2.36$.}
\label{fig:orderparameter}
\end{figure*}

Figure~\ref{fig:orderparameter} shows that the order parameter $S$, which represents
the degree of aggregation and is closely related to the weight average of vortex
mass, grows with increasing $\rho_0$ and $D_0$.
Interestingly, in the stationary state of the flagella system, $S$ increases
linearly with flagellum density when $\rho_0D_0^2 \leq 3$, which is reminiscent of
the linear relation between $S$ and $\rho_0D_0^2$ for $t/T_0\le 100$
in circle SPPs systems (Fig.~\ref{fig:particle_order_time}b). However, in circle
SPPs systems, $S$ continues to increase after $t/T_0=100$. We attribute the
linear relation of $S$ and $\rho_0$ in the flagella system to the effect of
volume exclusion.
The linear dependence of $S$ on the flagellum density agrees very well with the experimental
observations for sea-urchin sperm.
On the other hand, $S$ is found to increase as $S\sim D_0^3$, see
Fig.~\ref{fig:orderparameter}b, indicating the importance of a second length
scale, which should be related to the flagellum length.

\subsection{Discussion}

The effect of volume exclusion and synchronization of the flagellar beat manifests
itself in the following three main aspects.

First, as shown in Fig.~\ref{fig:snapshot}bc, the flagellum vortices are closed
rings composed of synchronized flagella. This closed structure makes the fusion
of large vortices to happen very infrequently, because the necessary force to open
such a structure to fuse two neighboring vortices is very large. Therefore, as
suggested by our circle SPPs simulations, the flagellum system experiences
vortex formation and reorganization only during the early stages of structure
formation, corresponding to period I in the circle SPP system, which leads
to vortices described by an order parameter $S$ which depends linearly on the
particle density. The next step to raise the aggregation number, corresponding
to period III in the circle SPP system, which requires the fusion of heavy
vortices, is prevented by their nearly impenetrable closed structure.

Second, an extraordinary heavy vortex cannot exist for long, because the
maximum mass of a vortex is determined by
the balance of forces between flagella. For a flagellum swimming at a
distance $R$ from the center, the
prevalent trajectory curvature is $1/R$, while the preferred trajectory
curvature of a flagellum is $2/D_0$. When $R < D_0/2$, the flagellum at $R$
pushes outwards and exert an outward force on other flagella in outer layers;
similarly, when $R > D_0/2$, a flagellum at $R$ pushes inwards and exerts an
inward force on other flagella in inner layers.
When $R$ exceeds $D_0$, an instability should develop, which leads to
a breakup into smaller vortices of radius $D_0/2$.
For systems with larger $D_0$, the region $R < D_0/2$ is larger and the
typical difference between $1/R$ and $2/D_0$ is smaller, so that
the maximum vortex mass is larger than for systems with smaller $D_0$.
Therefore, $S$ increases with increasing $D_0$ for fixed $\rho_0$, as
shown in Fig.~\ref{fig:orderparameter}ab.

Third, consider now a large flagellum density, for which the system is already full of
vortices of similar mass, frequently colliding with each other (see movie S3 in
supplemental material \cite{movie}). A further increase of $\rho_0$ starts to destroy
the vortex structure and the order parameter $S$ decreases,
as shown in Fig.~\ref{fig:orderparameter}a for $\rho_0D_0^2>3$.
Such a decrease of $S$ was not observed in the sea-urchin sperm experiment
\cite{Riedel2005} because the
experimental system was not strictly two-dimensional. We conjecture that at a
certain surface density of sperm, the substrate is completely packed with vortices
and cannot absorb any more cells. The local hexagonal order of the sperm
vortex array \cite{Riedel2005} is a clue for this close packing. A further
increase of the surface density of sperm
is not possible because higher density will cause more frequent collision and
consequently expel some sperm from the near-substrate layer.
It seems not possible \cite{Riedel_comm_2013} to obtain a higher surface density
of sperm than 6000/mm$^2$, the largest density investigated in
Ref.~\onlinecite{Riedel2005}. For higher densities in the experiments,
multiple layers of sperm developed, and the layers on top were neither
ordered nor destroying the pattern below \cite{Riedel_comm_2013}.
A possibility to increase the density further in experiments might be to
restrict the sperm in a narrow slit of one layer thickness between two flat
substrates.

The variance $\sigma$ of the frequency distribution also influence the order
parameter, as shown in Fig.~\ref{fig:orderparameter}b. $S$ is not sensitive
to $\sigma$ when $\sigma\leq 5\%$, but decreases with increasing $\sigma$
for $\sigma>5\%$. When there are large differences between the frequencies of
flagella in a vortex, the collisions between the undulating shapes increase
the short-range repulsion, separate the
flagella and cause a looser vortex structure, so that vortex break up more
easily in collisions with other vortices. At small $\sigma<5\%$, this
effect is small. Note that even for $\sigma=30\%$, vortices still exist, as
indicated by the density correlation functions. However, stable structures
hardly exist for a long time. The frequent fission and fusion of vortices
make $S$ small, although the system is not completely disordered.

\section{Summary \& Conclusions}
\label{sec:summary}

We have simulated systems of self-propelled particles with preferred circular trajectories
(circle SPPs) interacting via a velocity-trajectory coordination rule, and systems
of curved flagella propelled by a sinusoidal beating motion.
In both systems, we observe the formation of vortex arrays, controlled by
particle density, interaction range, and diameter of the preferred circular trajectory.

For the circle-SPP systems, the vortex array shows liquid-like rather than hexagonal
spatial order.  The diameter of the
vortices, $D_v$, is about the diameter of a single particle trajectory $D_0$
with a slight increase with the diameter of the interaction region $d$, but
is not sensitive to the particle density $\rho_0$.
The average distance between neighbor vortices, $D_{vv}$, is also not sensitive
to $\rho_0$, but increases quickly with increasing $d$. A transition from
a light-vortex-dominated state (at low $\rho_0$ and small $d$) to a
heavy-vortex-dominated state (at high $\rho_0$ or large $d$) is observed.

We use an order parameter $S$ to characterize the degree of the vortex formation.
By comparing the time evolution of $S$,
we find that the vortex formation can be divided into three time periods.
During period I, the particles collect neighbor ones
to form vortices and $S$ increase quickly with time. The increase of
$S(t)$ during period I is slightly elevated with $\rho_0$, but is not sensitive
to $d$. In the subsequent period II, $S(t)$ increases very slowly.  In period III, the
vortex mass increases again more rapidly through vortex
collision and fusion. Note that environmental noise is not described in
our model. Therefore the fission, fusion and displacement of vortices is
purely the result of multi-particle interactions of circle SPPs.

In order to compare with the experiments of sperm cells near surfaces, we have also
studied a more detailed model of curved, sinusoidally beating flagella. Vortex patterns
in this system emerge from the hard-core repulsion of the curved body of
the elongated self-propelled particles moving in a viscous environment.
In the simulations with anisotropic frictions, the collective motion of the
curved flagella system agrees very well with the behavior of the circle SPPs
system, as well as the phenomenon observed in the sea-urchin sperm
experiments \cite{Riedel2005}. As in the circle SPPs system, the average
size of the vortices $D_v$ equals approximately $D_0$, and slightly increases
with the flagellum density. By comparing $D_{vv}$ with the circle SPPs systems,
we find that the size of the effective interaction region of a curved
flagellum can be approximately identified with $0.9\sim1.0D_0$. The order
parameter $S$ increases with $\rho_0$ as well as $D_0$. The fraction
$D_{vv}/D_v=1.8\pm0.2$ coincides the value $1.86\pm0.52$ calculated by using
the data from \cite{Riedel2005} for sea-urchin sperm system.

In conclusion, the collective motion of self-propelled particles, which leads to
the formation of vortex arrays, can be well reproduced
by circle SPPs with a velocity-trajectory coupling interaction.
The velocity-trajectory coordination rule is a different interaction type than
the velocity coordination rules employed since the Vicsek model \cite{Vicsek1995}
for the simulations of collective motion.
Such an interaction mimics, for example, the hard-core interaction of curved,
sinusoidal beating flagella.
The analysis of a more specific model of beating flagella
allows to elucidate the features related to an explicit propulsion mechanism
and physical interactions.

\section{Acknowledgments}
Y.Y. gratefully acknowledges support from the National Natural Science
Foundation of China (Grants 21304020) and the
Shanghai Postdoctoral Scientific Program (Program No.11R21411200).
G.G. gratefully acknowledges support from the VW Foundation
(VolkswagenStiftung) through the program {\it Computer Simulation of
Molecular and Cellular Bio-Systems as well as Complex Soft Matter}.



\begin{thebibliography}{99}
\bibitem{parameters_flagellum}
For the flagellum model, we chose the parameters $l_0=0.5a$, $N_f=100$, monomer mass $m_f=10m$,
$k=2\times10^5k_BT/a^2$, $\kappa=10^4k_BT/a^2$, $A=0.2/a$, and $q=4\pi/N_f l_0$.
Volume exclusion is obtained by a repulsive Lennard-Jones potential with $\epsilon=15k_BT$ and
$r_0=a$.
\bibitem{parameters_AF}
For Brownian Dynamics simulations with anisotropic friction (AF), the
friction coefficients are chosen to be $\gamma_{\perp}=10.0$ and $\gamma_{\parallel}=16.8$, in
order to reproduce the single-flagellum velocity $v_{f,0}/(L_{fl}f_0)\approx0.055$ in MPC fluid when
$c_0L_{fl}=2$.
\end{thebibliography}
\end{document}